\documentclass{pasa}%

\usepackage{soul} 

\newcommand{\sigm}{\sigma_{\text{m}}}
\newcommand{\sigmu}{\sigma_{\text{m,uni}}}
\newcommand{\sigmreg}{\sigma_{\text{m},v_\text{g}=0}}
\newcommand{\sigmureg}{\sigma_{\text{m,uni},v_\text{g}=0}}

\title [Resolved gas kinematics in high SFR galaxies]{Resolved gas kinematics in a sample of low--redshift high star--formation rate galaxies}
\author[Mathew Varidel et al.]{Mathew Varidel$^{1,2}$, Michael Pracy$^1$,  Scott Croom$^{1,2}$,  Matt S. Owers$^{3,4}$ and Elaine Sadler$^{1,2}$\\
\affil{$^1$Sydney Institute for Astronomy, School of Physics, University of Sydney, NSW 2006, Australia}
\affil{$^2$ARC Centre of Excellence for All-Sky Astrophysics (CAASTRO)}
\affil{$^3$Department of Physics and Astronomy, Macquarie University, NSW 2109}
\affil{$^4$Australian Astronomical Observatory, P.O. Box 915, North Ryde, NSW 1670, Australia}
}

\jid{PASA}
\doi{10.1017/pas.\the\year.xxx}
\jyear{\the\year}

\usepackage[authoryear]{natbib}

\bibpunct{(}{)}{;}{a}{}{,}
\begin{document}%
\begin{abstract}
We have used integral field spectroscopy of a sample of six nearby ($z \sim 0.01 - 0.04$) high star-formation rate ($\text{SFR} \sim 10 - 40$ $\text{M}_\odot \text{/yr}$) galaxies to investigate the relationship between local velocity dispersion and star formation rate on sub-galactic scales. The low redshift mitigates, to some extent, the effect of beam smearing which artificially inflates the measured dispersion as it combines regions with different line-of-sight velocities into a single spatial pixel. We compare the parametric maps of the velocity dispersion with the H$\alpha$ flux (a proxy for local star-formation rate), and the velocity gradient (a proxy for the local effect of beam smearing). We find, even for these very nearby galaxies, the H$\alpha$ velocity dispersion correlates more strongly with velocity gradient than with H$\alpha$ flux -- implying that beam smearing is still having a significant effect on the velocity dispersion measurements. We obtain a first-order {\it non parametric} correction for the unweighted and flux weighted mean velocity dispersion by fitting a 2D linear regression model to the spaxel-by-spaxel data where the velocity gradient and the H$\alpha$ flux are the independent variables and the velocity dispersion is the dependent variable; and then extrapolating  to zero velocity gradient.  The corrected velocity dispersions are a factor of  $\sim  1.3 - 4.5$ and $\sim 1.3 - 2.7$ lower than the uncorrected flux-weighted and unweighted mean line-of-sight velocity dispersion values,  respectively. These corrections are larger than has been previously cited using disc models of the velocity and velocity dispersion field to correct for beam smearing.
The corrected  flux-weighted velocity dispersion values are $\sigma_m \sim 20 - 50$ km/s.


\end{abstract}
\maketitle%

\section{INTRODUCTION }
\label{sec:intro}
The kinematics of the gas component of galaxies is pivotal to understanding  their properties and evolutionary state at all cosmic epochs. In particular, the spatially-resolved line-of-sight gas-phase velocity dispersion (which is related to the local turbulence in the disc) obtained via integral field spectroscopy is crucial. Integral field observations of $z > 1$ galaxies have found high line-of-sight velocity dispersions, a factor of $\sim 4$--10 times larger than those of local galaxies \citep[e.g.][]{genzel2006,law2007,forster-schreiber2009,wright2009,Maiolino2010,epinat2010,lemoine-busserolle2010,gnerucci2011,vergani2012,newman2013,wisnioski2015}.  
These high line-of-sight velocity dispersions suggest a turbulent
inter stellar medium. This turbulence provides support against gravitational 
collapse and implies that star-formation should take place in large
clumps which are massive enough to collapse out of this turbulent
medium \citep{elmegreen2009}. This provides an explanation for the regularly rotating but
photometrically  irregular (clumpy) galaxies common at $z>1$ \citep[e.g.][]{wisnioski2011}.

The physical mechanism(s) causing and preserving these high line-of-sight
velocity dispersions remains an open question. Several possibilites
have been suggested, these include gravitational instability
\citep[e.g.][]{bournaud2010} or generation during the initial
gravitational collapse \citep{elmegreen2010}. In principle, these high velocity dispersions could be caused by feedback from star-formation in the disc. However, \citet{genzel2011} found only a weak correlation between star-formation and line-of-sight velocity dispersion. The high velocity dispersions are often interpreted as being consistent with, and evidence for, cold flow accretion from the inter--galactic medium at high redshift \citep{aumer2010}. 

In contrast, observations by \citet{green2010,green2014} of a sample of  strongly star-forming galaxies at z$\sim$0.05--0.3 found that the spatially resolved line-of-sight  velocity dispersion, measured from the H$\alpha$ emission line, was similar to the high dispersions observed at $z >1$. By combining samples spanning a range of star-formation rates and redshifts \citep{law2009,law2007,yang2008,garrido2002,epinat2008a,epinat2008b,epinat2009,lemoine-busserolle2010,cresci2009,contini2012,wisnioski2011,jones2010,swinbank2012,green2014}, \citet{green2010,green2014}
found that velocity dispersion did correlate with the star-formation rate. From this they suggested that star-formation is the energetic driver of galaxy disc turbulence at both high and low redshift.
In this picture, selection effects which bias toward the most star-forming galaxies at high redshift and the rarity of such objects at low redshift result in the difference in gas phase line-of-sight velocity dispersions observed at the different epochs. 

Since these measurements in general have spatial resolutions that correspond to physical scales that are significant with respect to the galaxy size, a key observational issue to be resolved is the effect of beam smearing which always  acts to increase the measured dispersion on such observations. \citet{davies2011} have demonstrated that biases, as a consequence of beam smearing effects, in such measurements can be severe. The sample of \citet{green2010}  have redshifts of $z \sim 0.05$--$0.3$, which  typically correspond to physical-scale resolutions of $\sim$1--4\,kpc (FWHM of the PSF). At this spatial resolution the smearing due to convolution with the PSF can cause fundamental problems with the interpretation of spatially resolved spectroscopy \citep[e.g.][]{davies2011,pracy2010}. \citet{green2014} addressed this issue by carefully modelling the contribution of beam smearing to the observed velocity dispersion on a per spaxel basis.  This technique assumes a disc model and an exponential light profile to calculate the contribution to the velocity width arising from the unresolved velocity gradient across a spatial resolution element. This correction will be valid when the disc model describes the velocity field well and the H$\alpha$ flux distribution
is exponential. However, the correction is not valid for non-disc galaxies or disc galaxies where the disc model does not predict the true `infinite resolution' velocity field accurately. If, for example, the inner unresolved velocity gradient is steeper than the model prediction or the true H$\alpha$ surface brightness is steeper than exponential then such a correction would under-predict the contribution from `beam smearing'. For the disc galaxies in \citet{green2014} they find a  median correction to the flux-weighted mean of the  spatially-resolved line-of-sight gas-phase velocity of just 3.6\,km\,s$^{-1}$, although in a few cases the correction is 
much more significant ($\sim$a factor of 2).

In this paper, we present results from integral field spectroscopic observations of a  small sample of six low redshift ($z<0.04$) and high star-formation rate  ($\text{SFR} \sim 10 - 40$ $\text{M}_\odot \text{/yr}$)
galaxies selected from the Sloan Digital Sky Survey  \citep[SDSS;][]{abazajian2009}. While galaxies with such high star-formation rates are rare in the local Universe the wide-area, and therefore large volume at low redshift, probed by the SDSS allows the selection of a small sample of such objects. The low redshift naturally provides a good physical-scale resolution and so minimizes the effects of beam smearing. In Section \ref{sec:sample} we present details of our sample selection, observations and data reduction. In Section \ref{sec:analysis} we outline our data analysis including emission line fitting and measuring  local velocity gradients from the data. In Section \ref{sec:results} we present our results, including the position of our sample galaxies in the velocity dispersion--star formation plane and a simple non-parametric approach to correcting the effects of beam smearing.  Throughout this paper we convert from observed to physical units
assuming a $\Omega_{M}=0.3$, $\Omega_{\Lambda}=0.7$ and
$H_{0}=70$\,km\,s$^{-1}$\,Mpc$^{-1}$ cosmology.

\section{Sample selection, observations and data reduction}
\label{sec:sample}
We required a sample of high star-formation rate and low redshift galaxies. We selected a sample of galaxies from the SDSS with total star formation rates above 20\,M$_{\odot}$ per year using the MPA/JHU value added galaxy catalogue. The star formation rates are calculated using the techniques outlined in \citet{brinchmann2004} and assume a \citet{kroupa2001} Initial Mass Function (IMF). We set a redshift limit of z = 0.04, which is a trade off between targeting galaxies as nearby as possible to gain high physical-scale resolution and having enough cosmic volume to define a reasonable number of target galaxies. 
At z = 0.04, 1 arcsecond corresponds to just 0.8 kpc which is an improvement of $\sim$2--4 over previous studies \citep[e.g.][]{green2010,green2014}.
The spectrum for every candidate was examined by eye and spurious objects removed. The resulting catalogue contains a total of 20 galaxies fulfilling the selection criteria. Six of these objects were
observable from Sidding Spring Observatory at the time of our observing run in April 2013. These six galaxies are the sample analysed in this paper. The targets have  stellar masses in the range $10.5 < \log({\rm stellar\, mass}) < 11.1$ \citep{kauffmann03}.

 A summary of our target galaxies is given in Table \ref{tab:targets}. In order to simplify comparison of the SFR from  \citet{brinchmann2004} with literature values and our own measurements we have converted their SFR estimates from a \citet{kroupa2001} to a  \citet{chabrier2003} IMF using the relationship given in \citet{madau2014}.

\begin{table*}
\caption{Observed galaxies.}
\label{tab:targets}
\begin{center}
\begin{tabular}{c c c c c c c cc}
\hline\hline
Label&Name&RA&Dec& $ z^a $  & log(stellar mass)  & $\text{SFR}^b$     &  $\text{SFR}^c$   \\
        &         &     &      &              &  M$_\odot$            & (M$_{\odot}$/yr)    &   (M$_{\odot}$/yr) \\
\hline
A & J112545.04+144035.6 & 11:25:45.04 & +14:40:35.67 & 0.0342&   $11.05_{-0.10}^{+0.09} $       & $25_{-8}^{+16} $ &  $16\pm 6 $   \\
B & J115705.93+010732.1 & 11:57:05.93 & +01:07:32.13 & 0.0395 &   $11.08_{-0.10}^{+0.09}$        & $23_{-8}^{+13} $ &  $31\pm 11 $  \\
C & J145129.30+092005.8 & 14:51:29.30 & +09:20:05.89 & 0.0294 &   $10.96_{-0.10}^{+0.10} $      & $16_{-10}^{+25} $  &  $34\pm 17 $  \\
D & J152429.54+082223.5 & 15:24:29.54 & +08:22:23.56 & 0.0363 &   $11.10_{-0.09}^{+0.10}$       & $15 _{-9}^{+23} $  & $20\pm 8 $   \\
E & J152527.48+050029.9 & 15:25:27.48 & +05:00:29.92 & 0.0358  &   $10.52_{-0.09}^{+0.09} $      & $16 _{-2}^{+2} $     &  $23\pm 6 $     \\
F & J153000.83+125921.5 & 15:30:00.83 & +12:59:21.56 & 0.0134 &    $10.91_{-1.0}^{+0.09}$        & $13_{-7}^{+19} $    &  $12\pm 1 $     \\
\hline\hline
\end{tabular}
\end{center}
\tabnote{\hspace{19.5mm}$^a$Redshifts obtained from the Sloan Digital Sky Server (SDSS).}
\tabnote{\hspace{19.5mm}$^b$Star-Formation Rate (SFR) obtained from \citet{brinchmann2004}  and converted to a  \citet{chabrier2003} IMF.} 
\tabnote{\hspace{19.5mm}$^c$Calculated from this work by summing over IFU field-of-view. Assumes \citet{chabrier2003} IMF. } 
\end{table*}

\begin{figure*}
\centering
\begin{tabular}{ c\x c\x c\x c\x c\x c\x c\x}

 & &
\hspace{5mm}\includegraphics[keepaspectratio=true,height=32mm,width=32mm,angle=270,trim=34mm 24mm 207mm 38mm,clip=true]{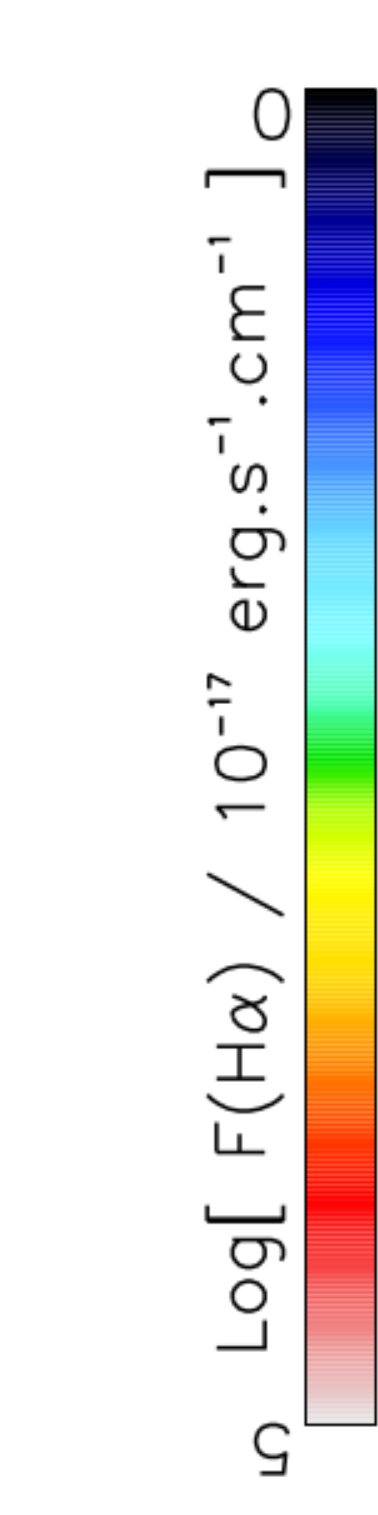} &
\hspace{5mm}\includegraphics[keepaspectratio=true,height=32mm,width=32mm,angle=270,trim=34mm 24mm 207mm 38mm,clip=true]{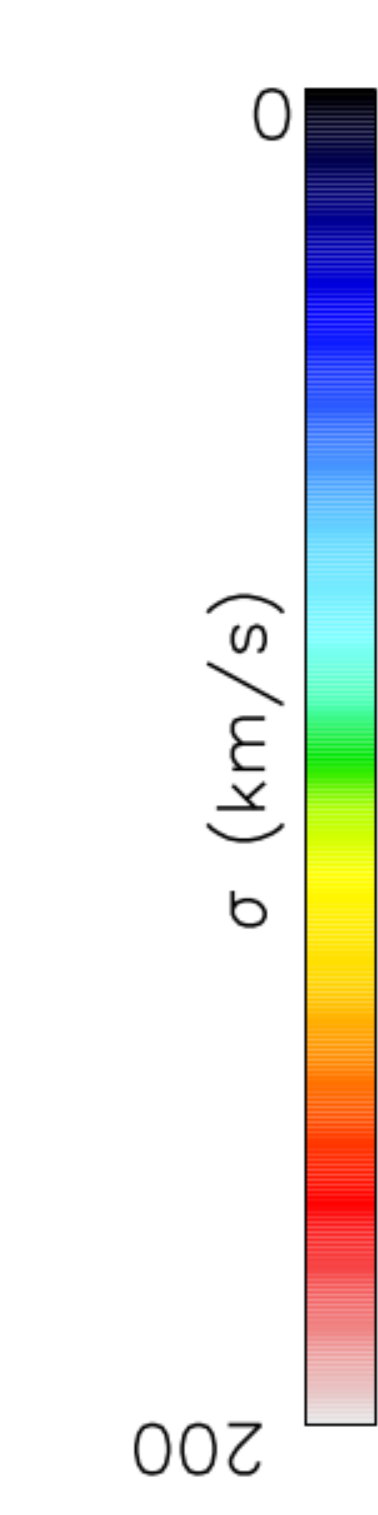} &
\hspace{5mm}\includegraphics[keepaspectratio=true,height=32mm,width=32mm,angle=270,trim=34mm 24mm 207mm 38mm,clip=true]{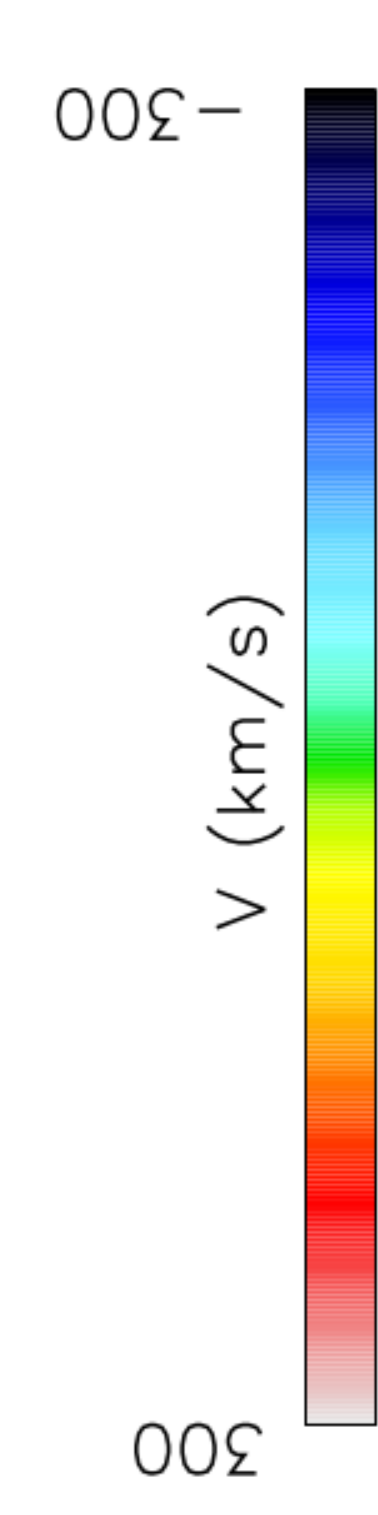} &
\hspace{5mm}\includegraphics[keepaspectratio=true,height=32mm,width=32mm,angle=270,trim=34mm 24mm 207mm 38mm,clip=true]{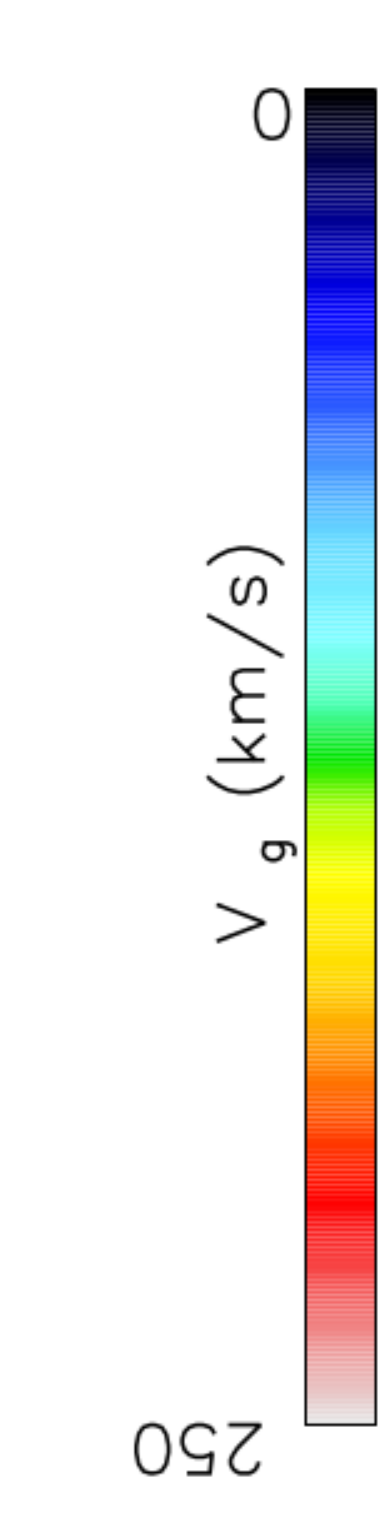} \\

A & \raisebox{\dimexpr-30.68mm+\ht\strutbox}{\includegraphics[keepaspectratio=true,height=27.1mm,width=27.5mm]{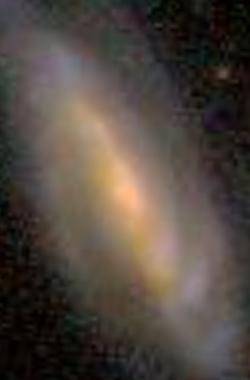}} &
\includegraphics[keepaspectratio=true,height=32mm,width=25mm,angle=180,trim=127mm 30mm 32mm 20mm,clip=true]{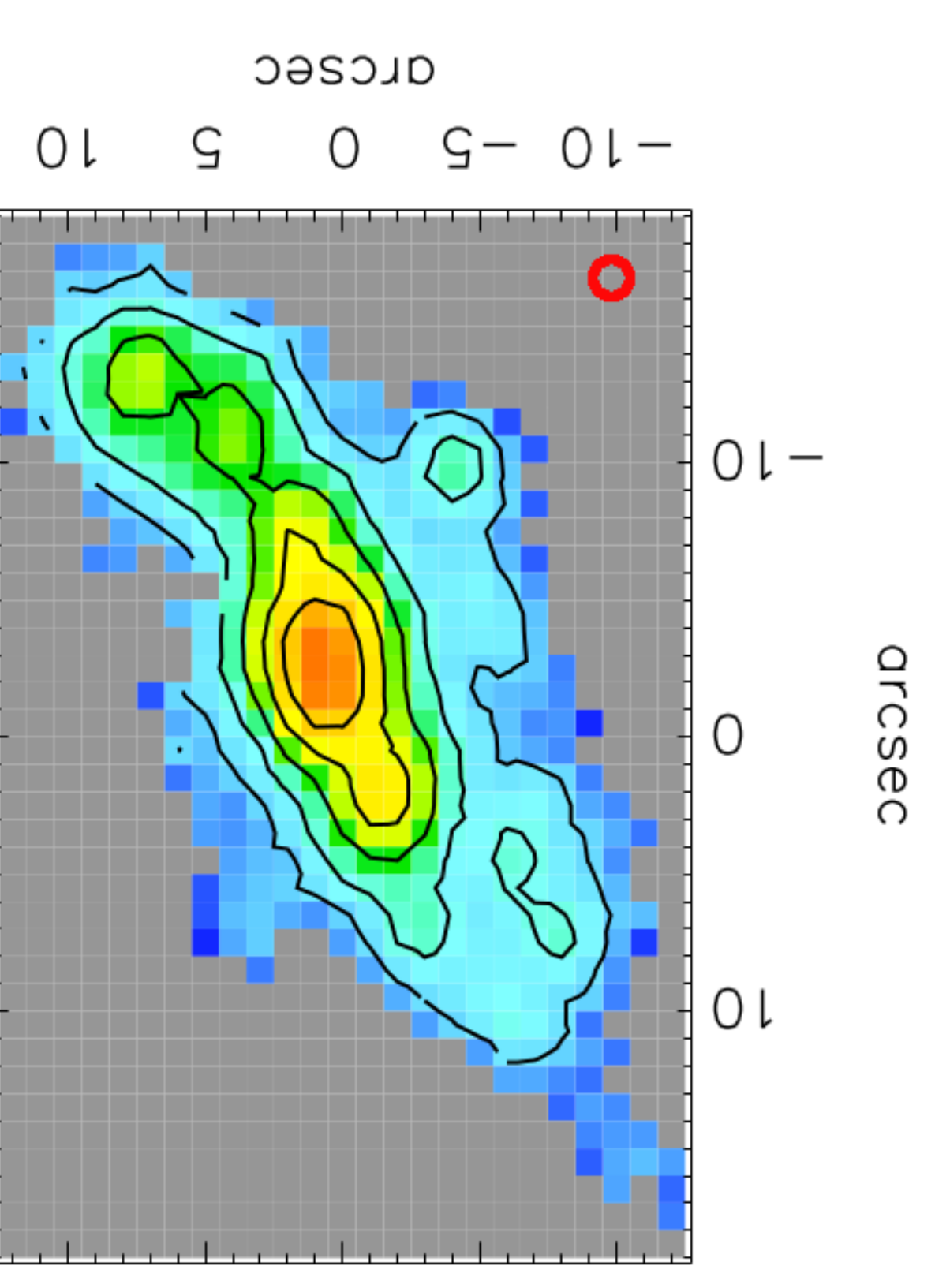} &
\includegraphics[keepaspectratio=true,height=32mm,width=25mm,angle=180,trim=127mm 30mm 32mm 20mm,clip=true]{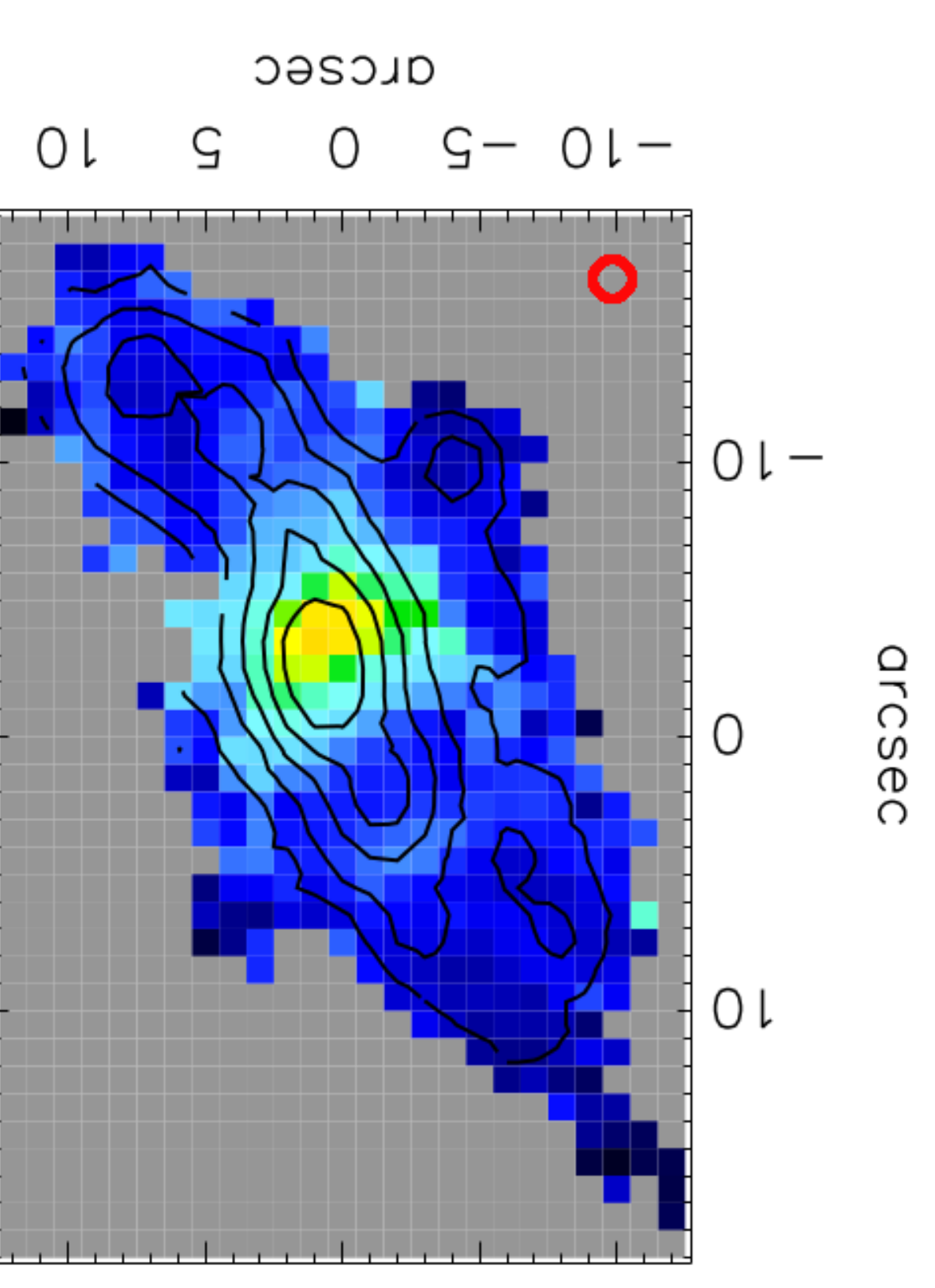} &
\includegraphics[keepaspectratio=true,height=32mm,width=25mm,angle=180,trim=127mm 30mm 32mm 20mm,clip=true]{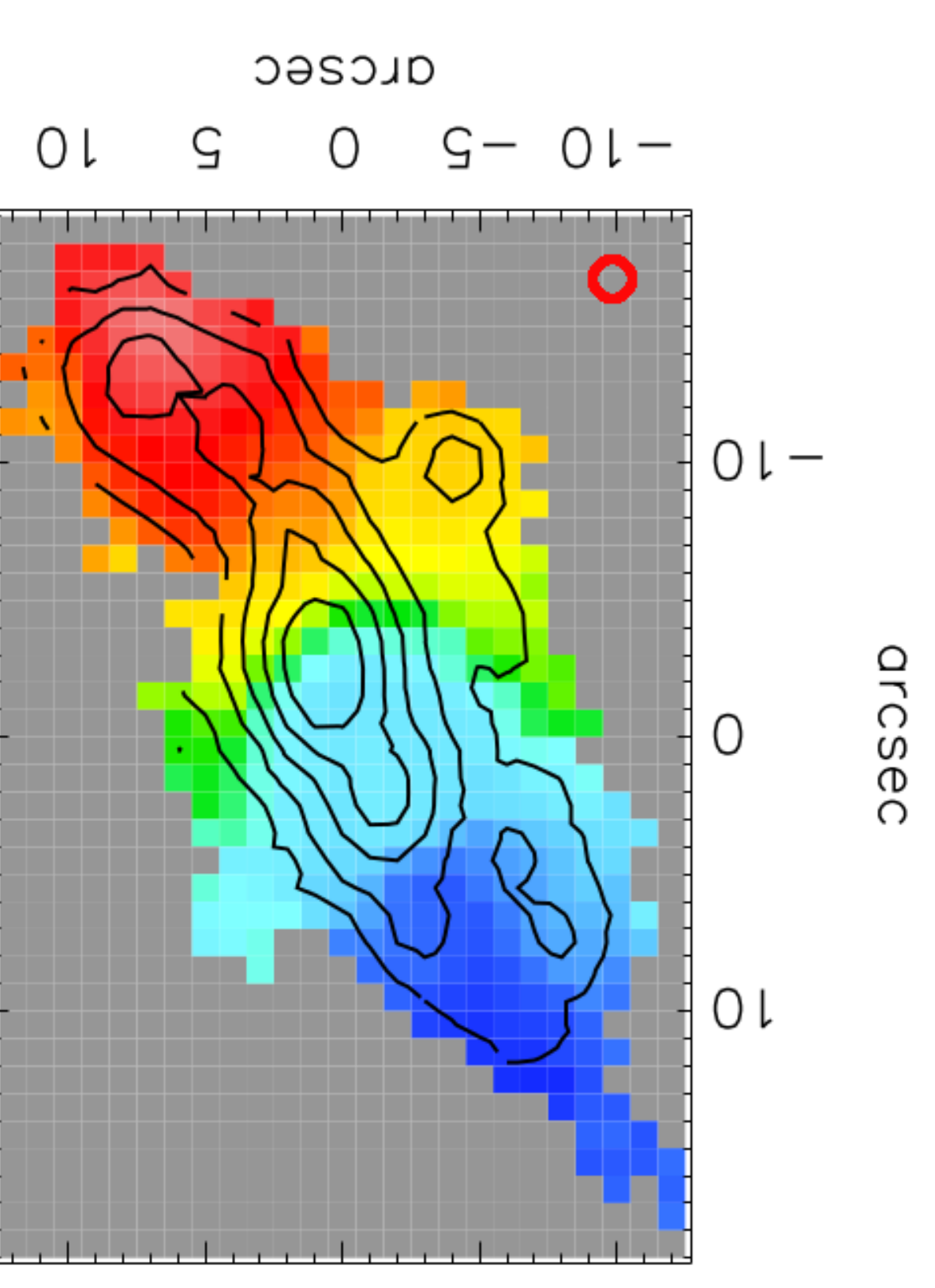} &
\includegraphics[keepaspectratio=true,height=32mm,width=25mm,angle=180,trim=127mm 30mm 32mm 20mm,clip=true]{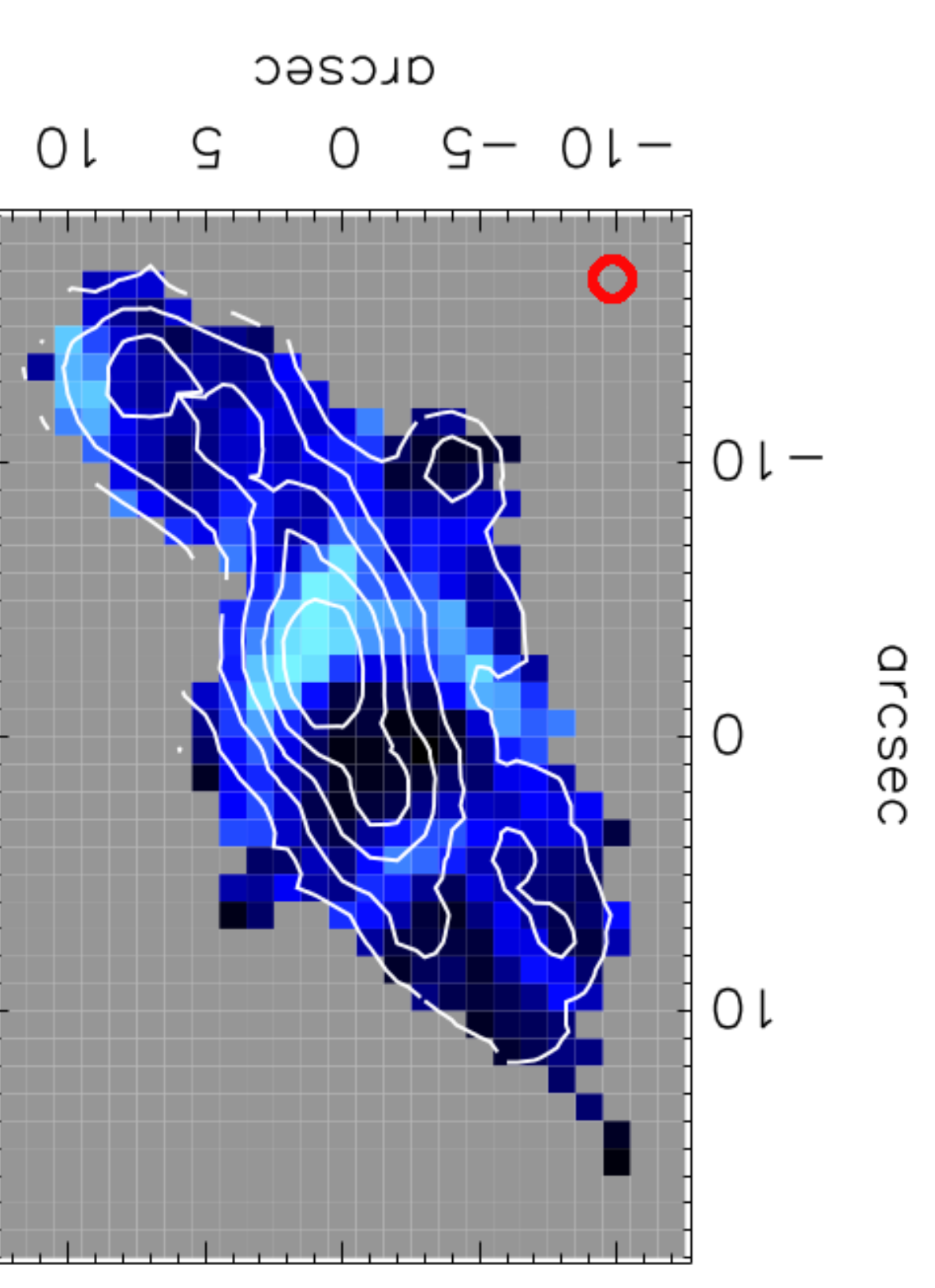} \\

B & \raisebox{\dimexpr-30.68mm+\ht\strutbox}{\includegraphics[keepaspectratio=true,height=27.1mm,width=27.5mm]{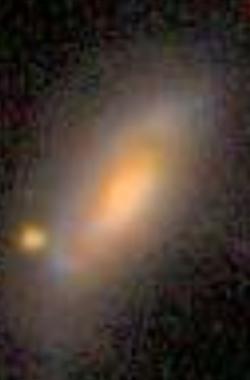}} &
\includegraphics[keepaspectratio=true,height=32mm,width=25mm,angle=180,trim=127mm 30mm 32mm 20mm,clip=true]{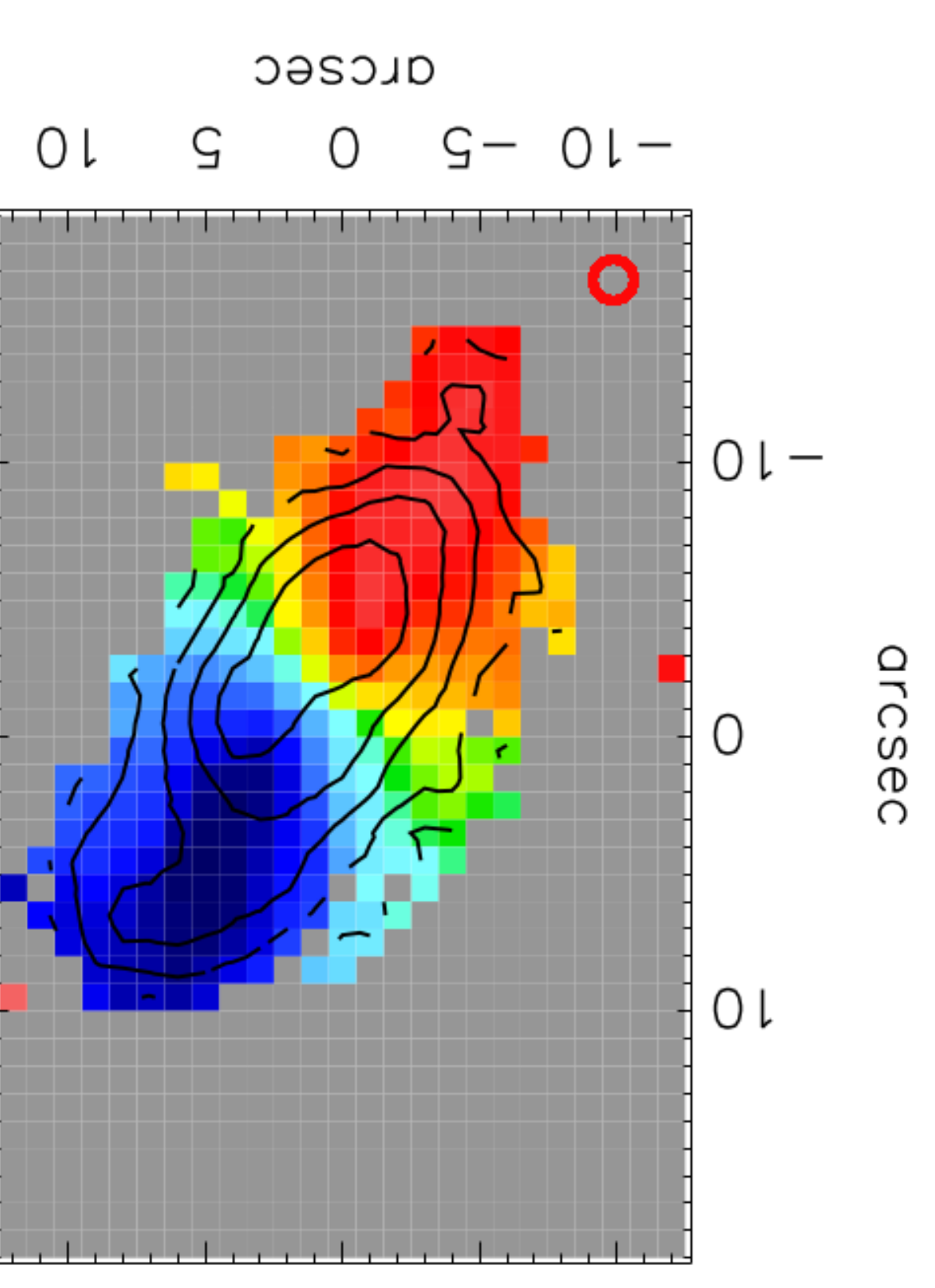} &
\includegraphics[keepaspectratio=true,height=32mm,width=25mm,angle=180,trim=127mm 30mm 32mm 20mm,clip=true]{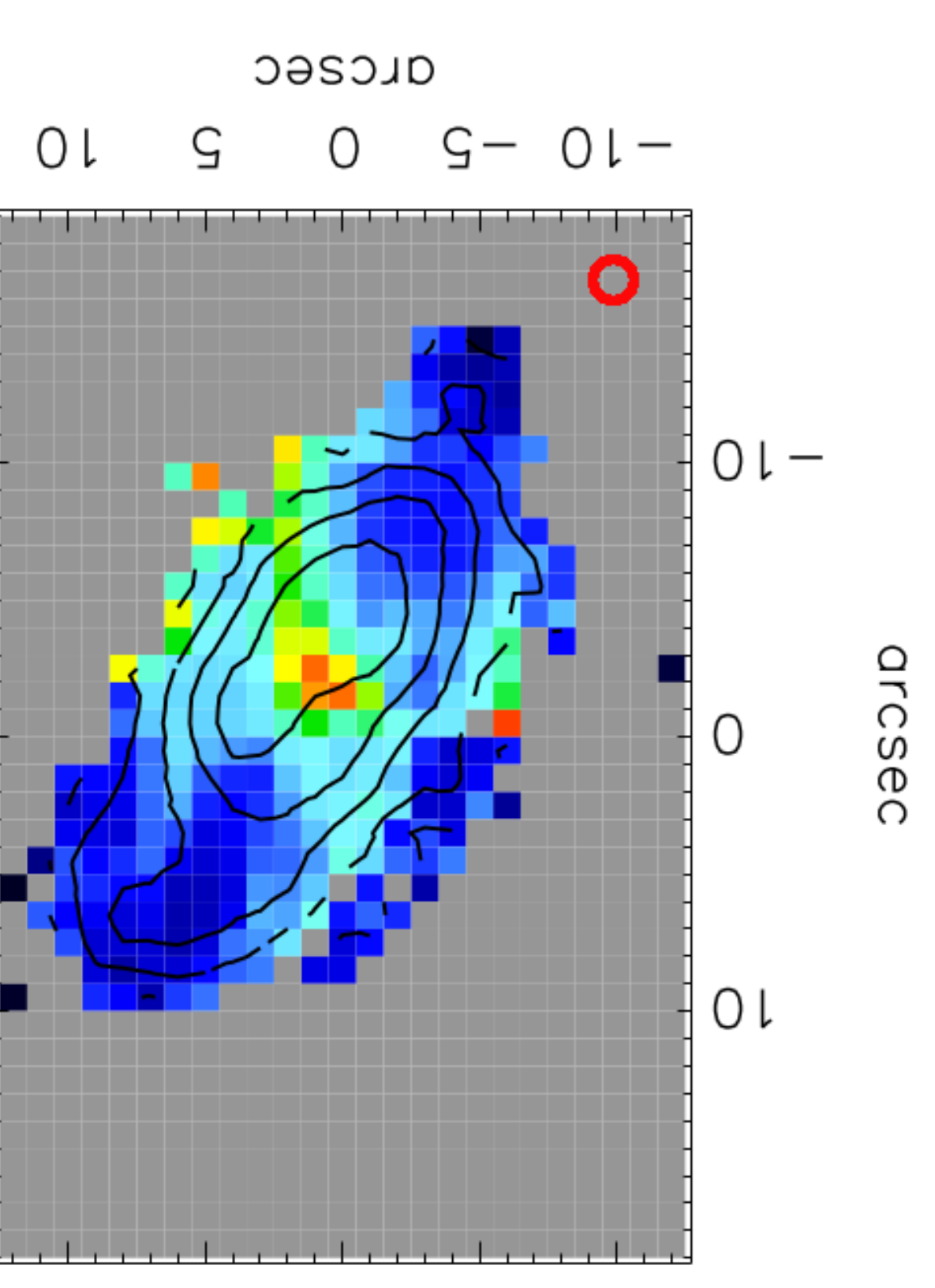} &
\includegraphics[keepaspectratio=true,height=32mm,width=25mm,angle=180,trim=127mm 30mm 32mm 20mm,clip=true]{"gal3_p1b"} &
\includegraphics[keepaspectratio=true,height=32mm,width=25mm,angle=180,trim=127mm 30mm 32mm 20mm,clip=true]{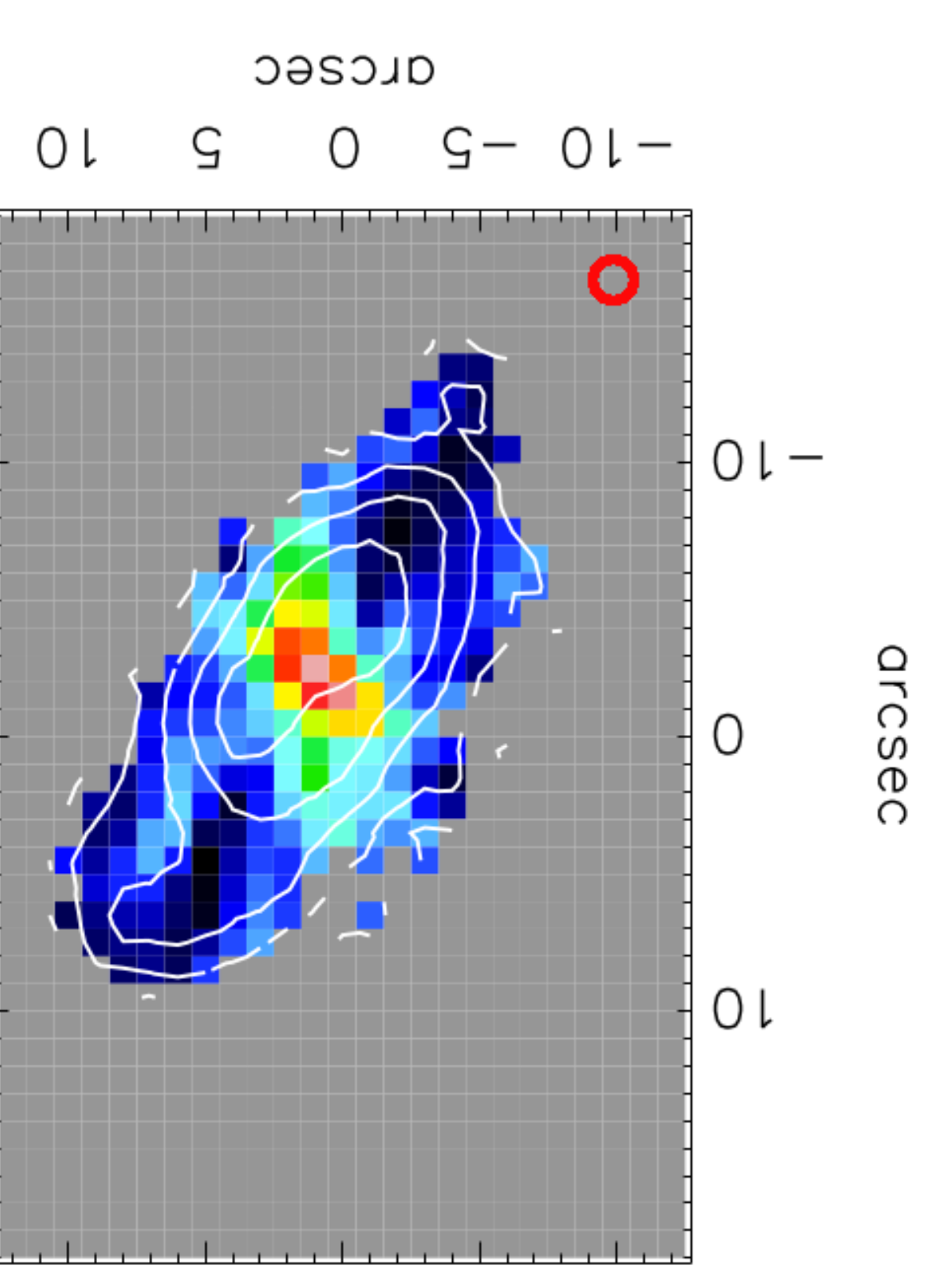} \\

C & \raisebox{\dimexpr-30.68mm+\ht\strutbox}{\includegraphics[keepaspectratio=true,height=27.1mm,width=27.5mm]{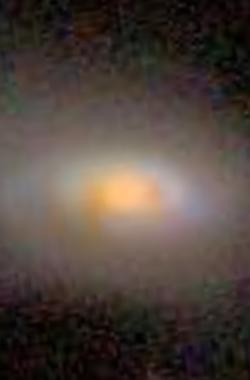}} &
\includegraphics[keepaspectratio=true,height=32mm,width=25mm,angle=180,trim=127mm 30mm 32mm 20mm,clip=true]{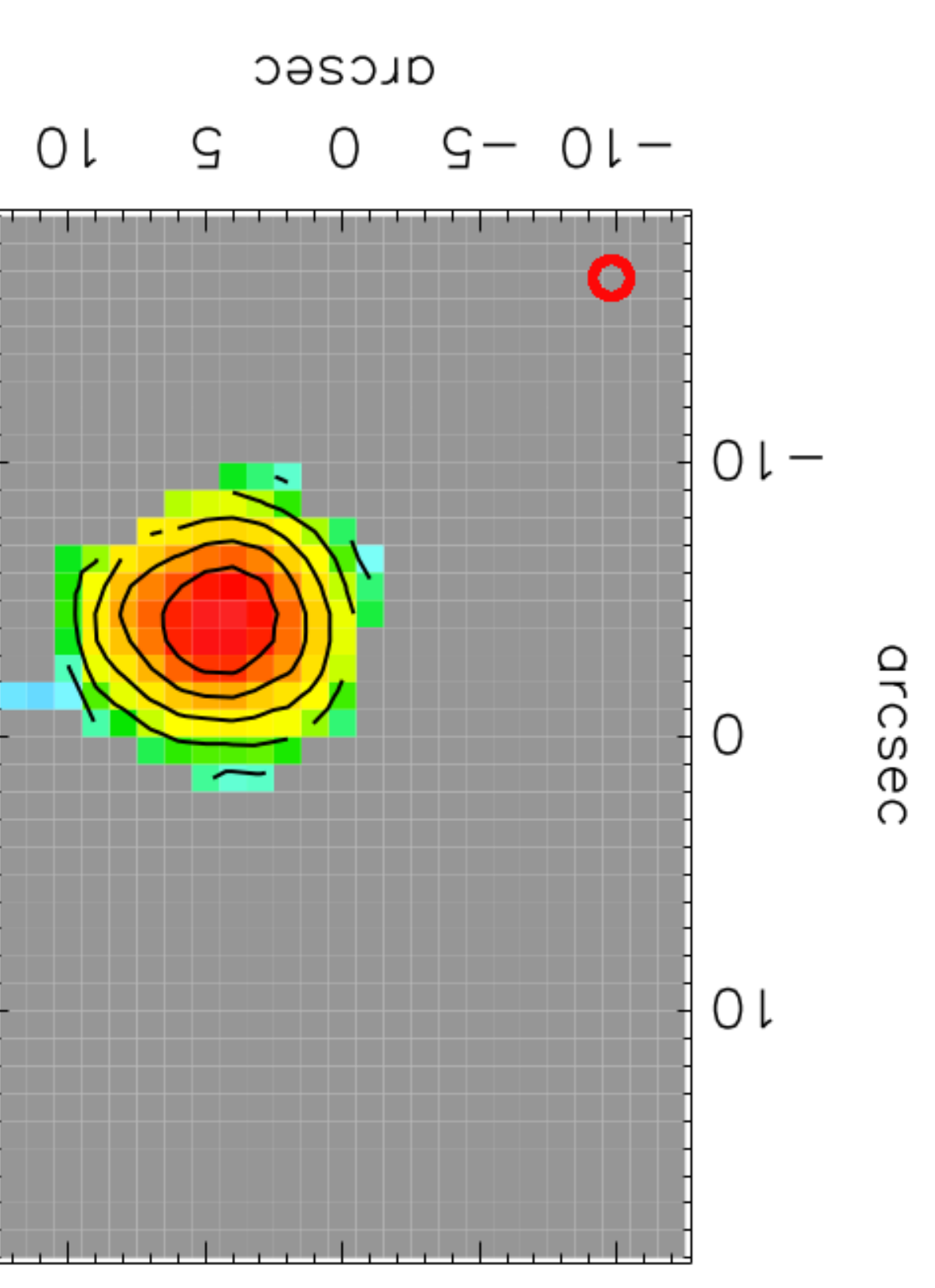} &
\includegraphics[keepaspectratio=true,height=32mm,width=25mm,angle=180,trim=127mm 30mm 32mm 20mm,clip=true]{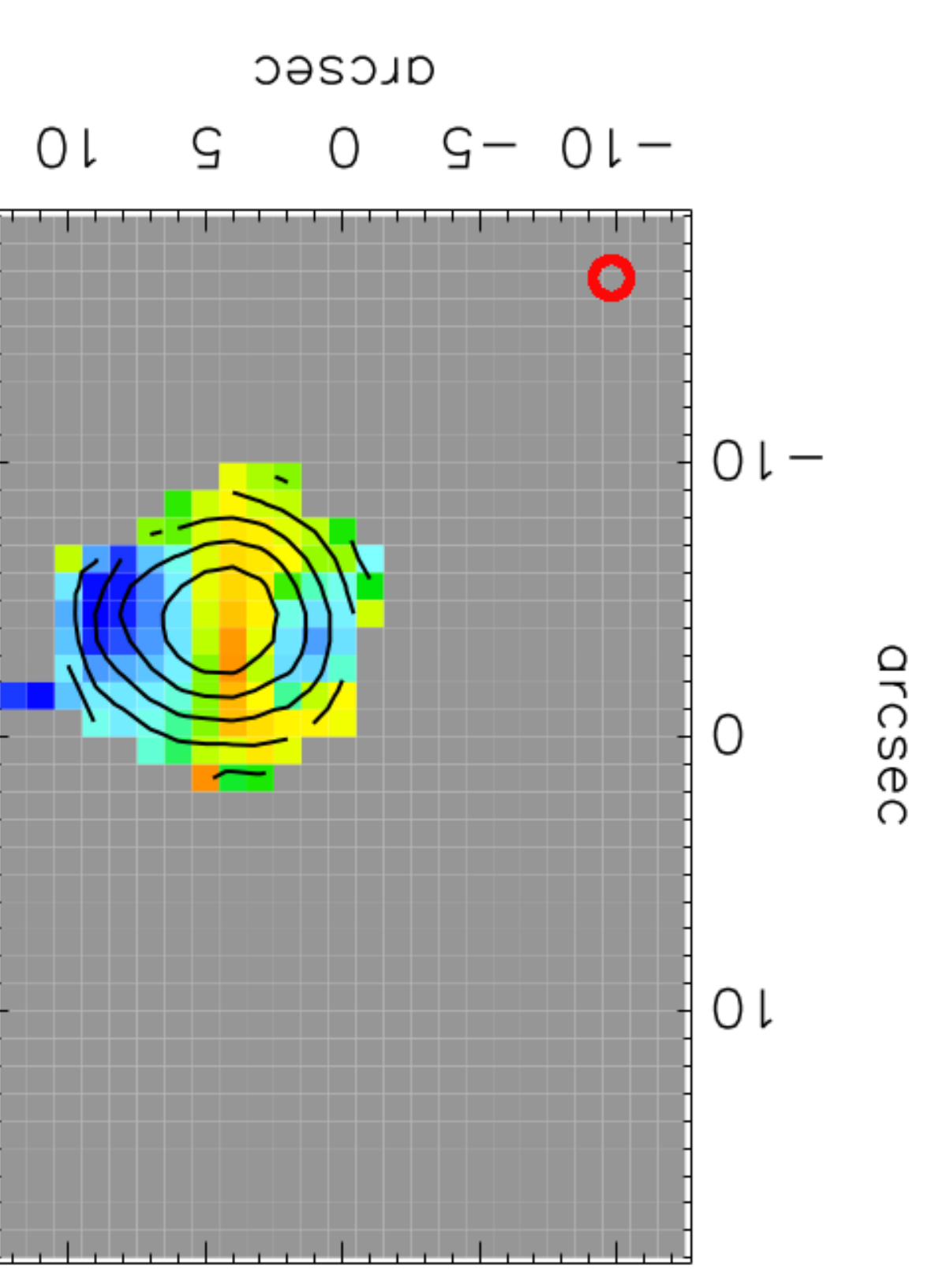} &
\includegraphics[keepaspectratio=true,height=32mm,width=25mm,angle=180,trim=127mm 30mm 32mm 20mm,clip=true]{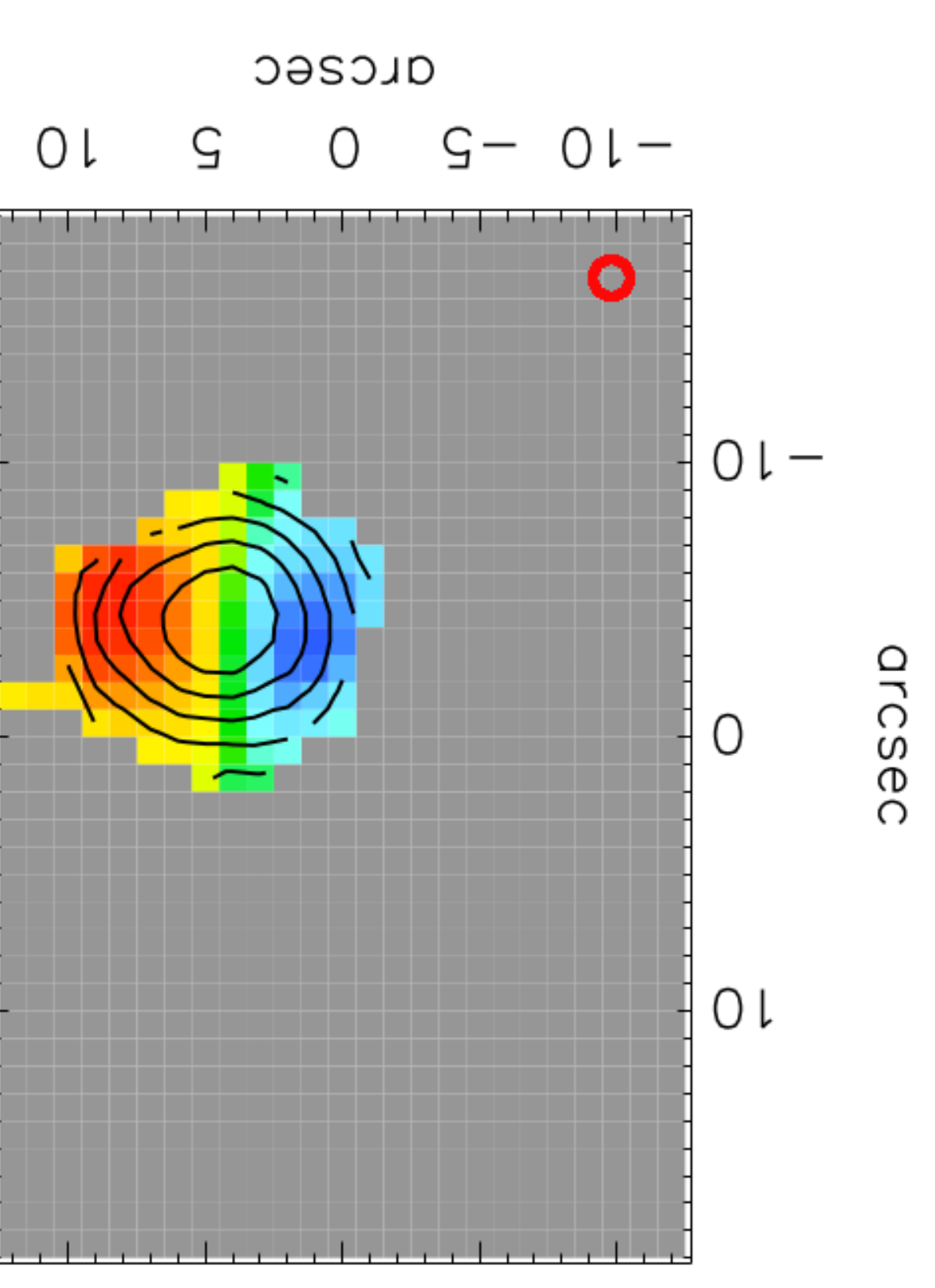} &
\includegraphics[keepaspectratio=true,height=32mm,width=25mm,angle=180,trim=127mm 30mm 32mm 20mm,clip=true]{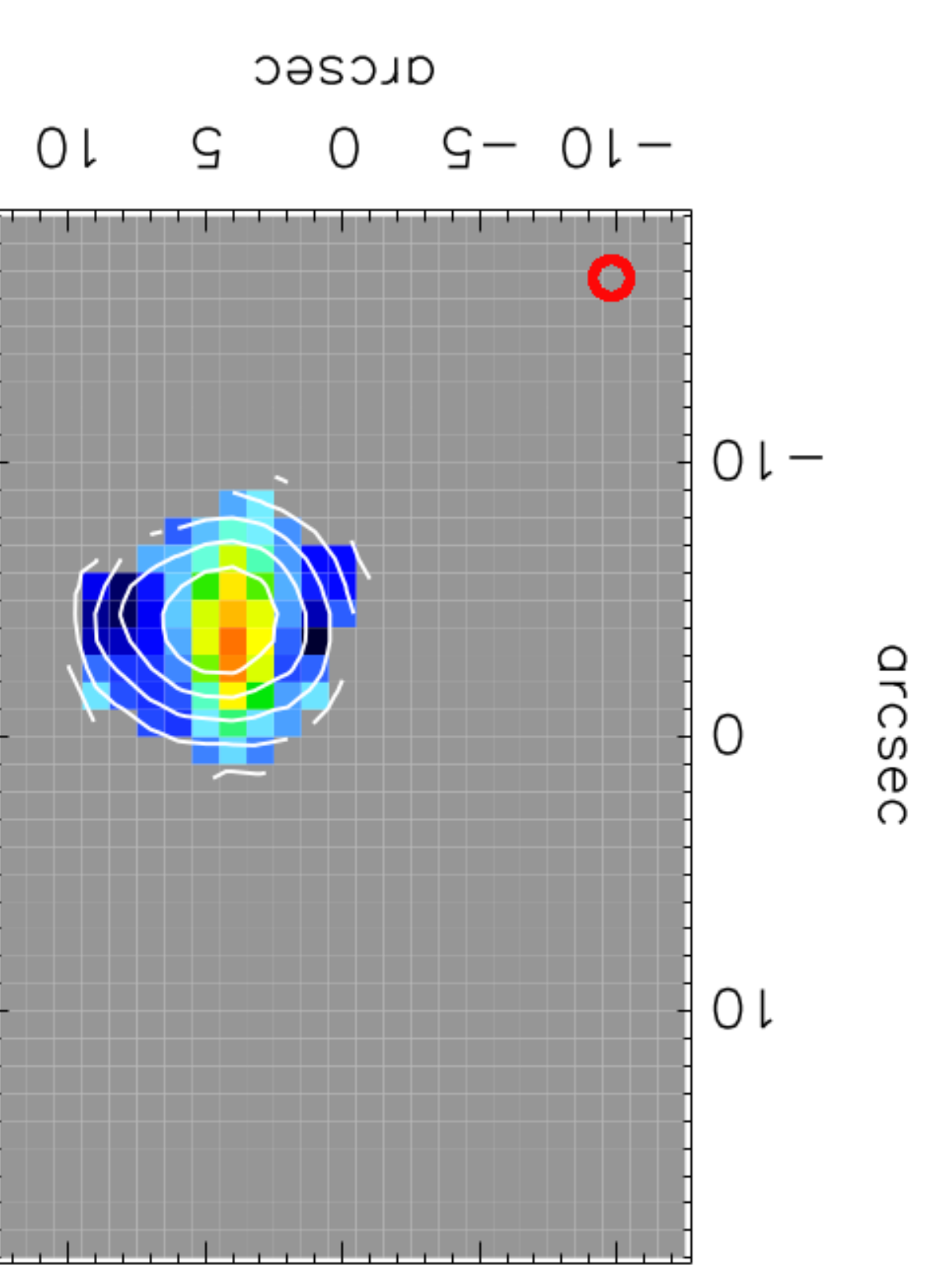} \\

D & \raisebox{\dimexpr-30.68mm+\ht\strutbox}{\includegraphics[keepaspectratio=true,height=27.1mm,width=27.5mm]{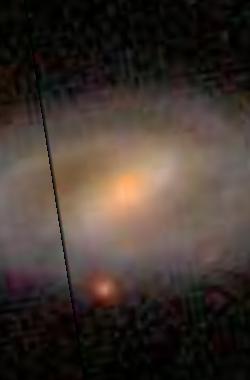}} &
\includegraphics[keepaspectratio=true,height=33mm,width=25mm,angle=180,trim=127mm 30mm 32mm 20mm,clip=true]{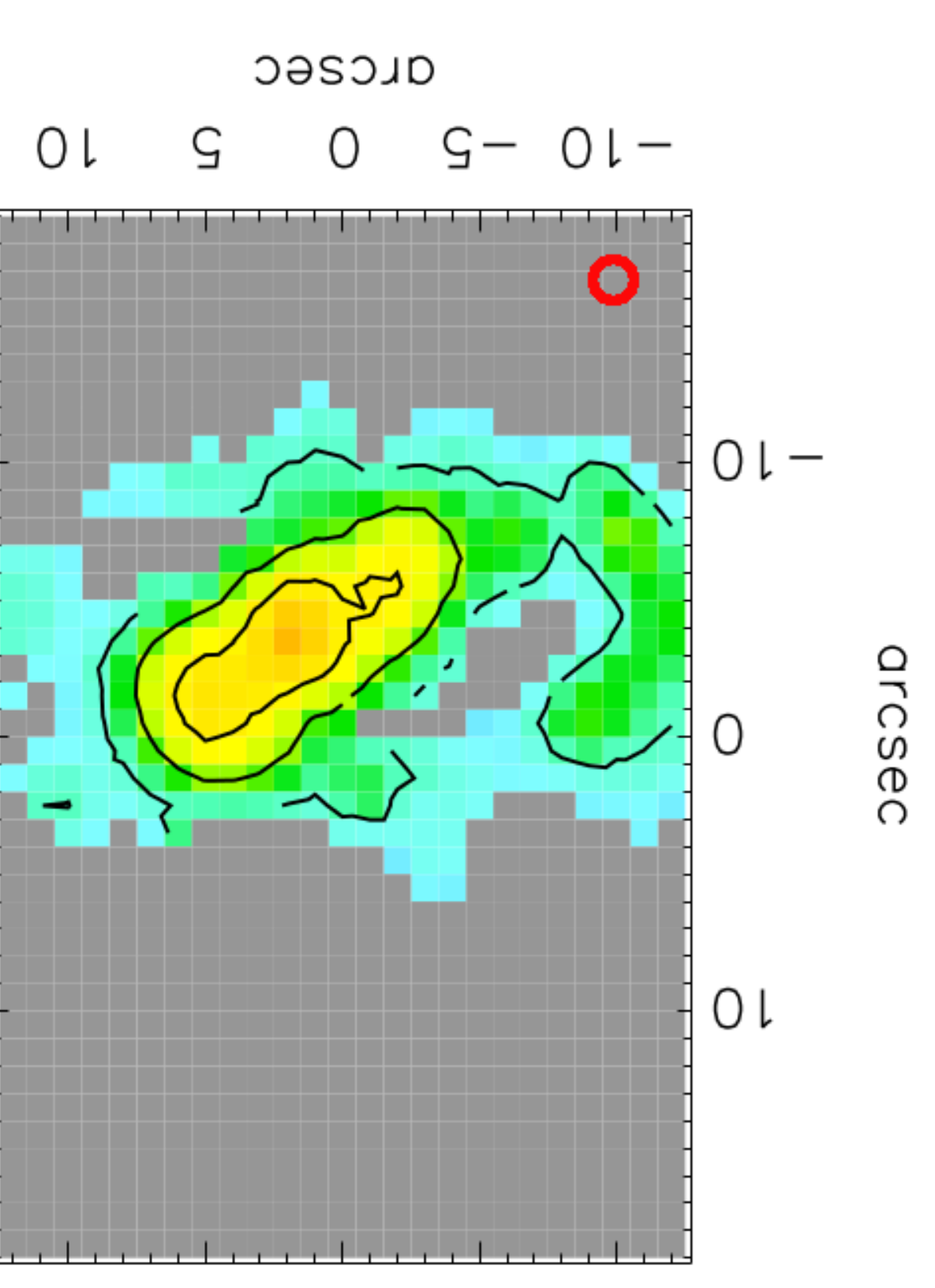} &
\includegraphics[keepaspectratio=true,height=33mm,width=25mm,angle=180,trim=127mm 30mm 32mm 20mm,clip=true]{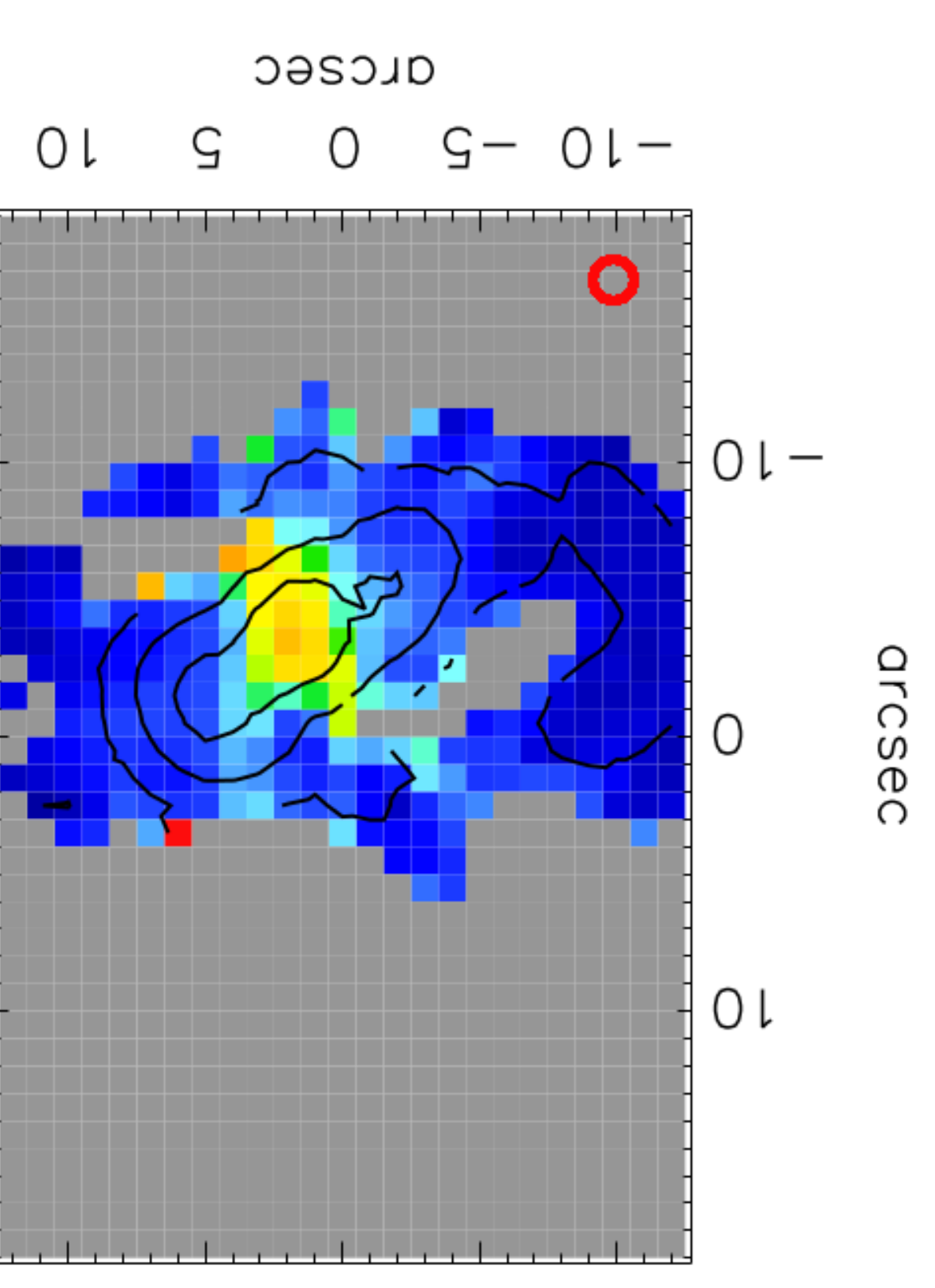} &
\includegraphics[keepaspectratio=true,height=33mm,width=25mm,angle=180,trim=127mm 30mm 32mm 20mm,clip=true]{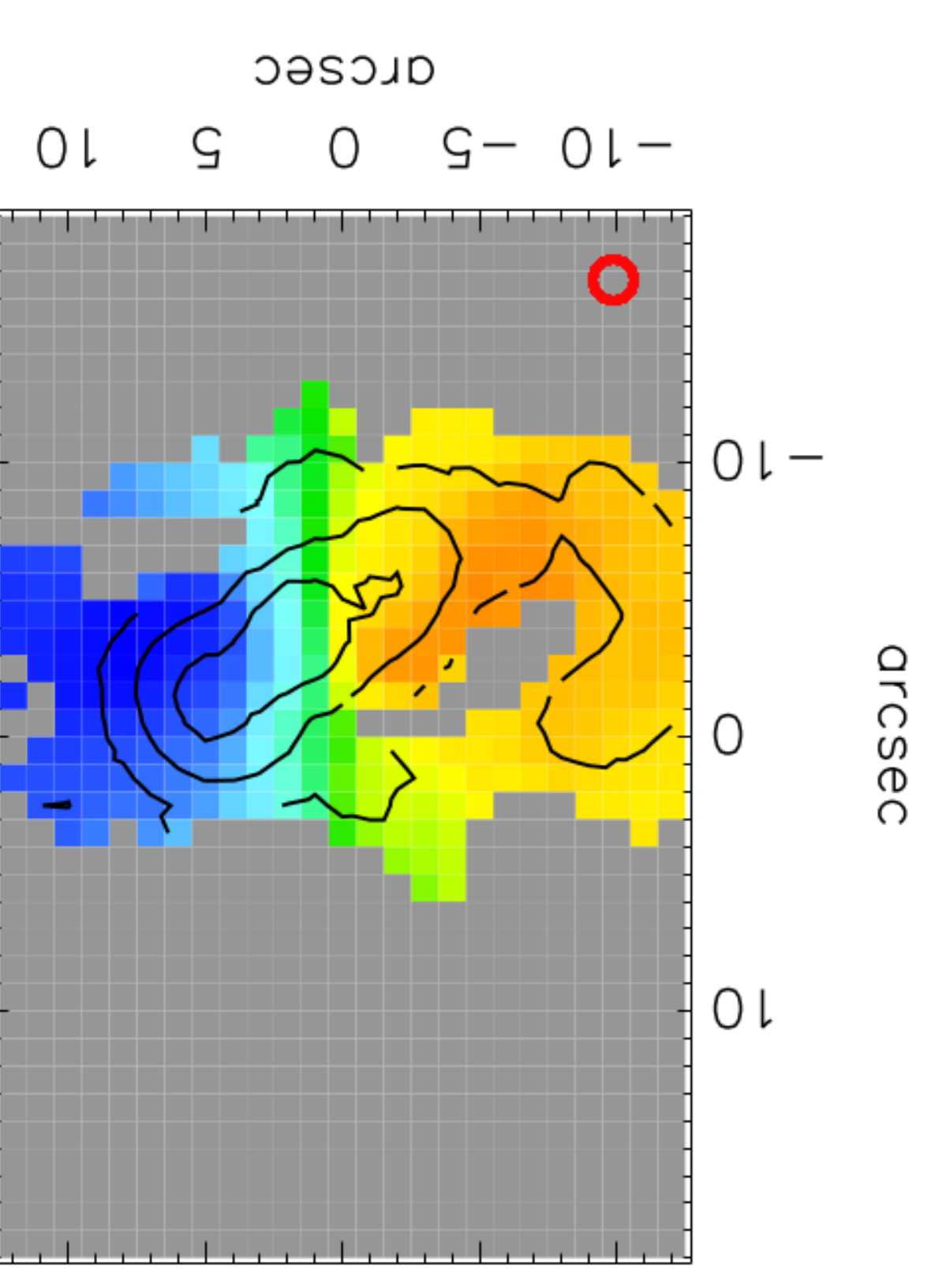} &
\includegraphics[keepaspectratio=true,height=33mm,width=25mm,angle=180,trim=127mm 30mm 32mm 20mm,clip=true]{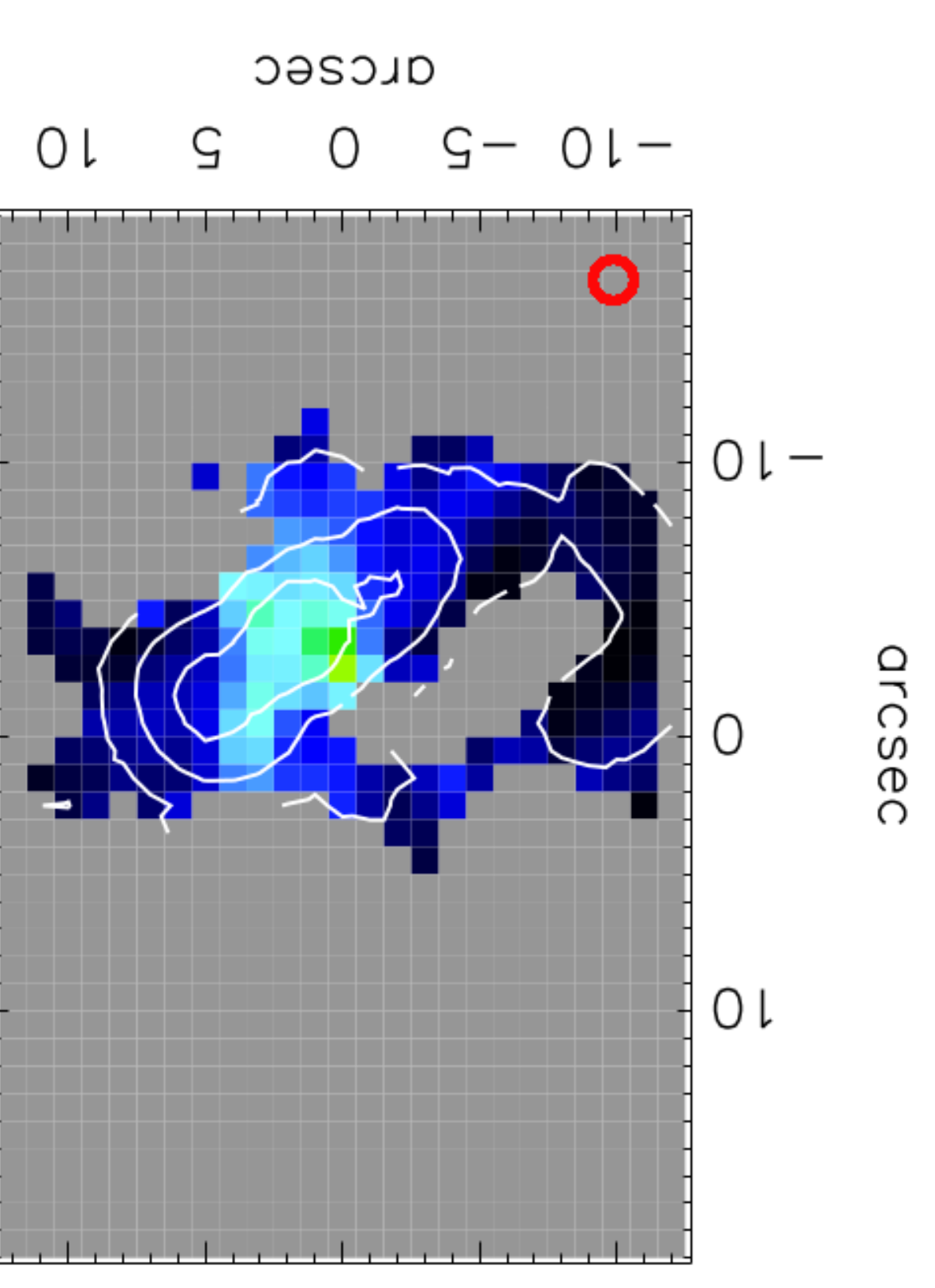} \\

E & \raisebox{\dimexpr-30.68mm+\ht\strutbox}{\includegraphics[keepaspectratio=true,height=27.1mm,width=27.5mm,trim=0mm 0mm 0mm 0mm]{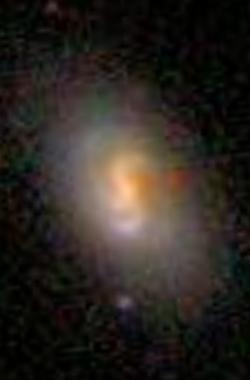}} &
\includegraphics[keepaspectratio=true,height=33mm,width=25mm,angle=180,trim=127mm 28mm 32mm 20mm,clip=true]{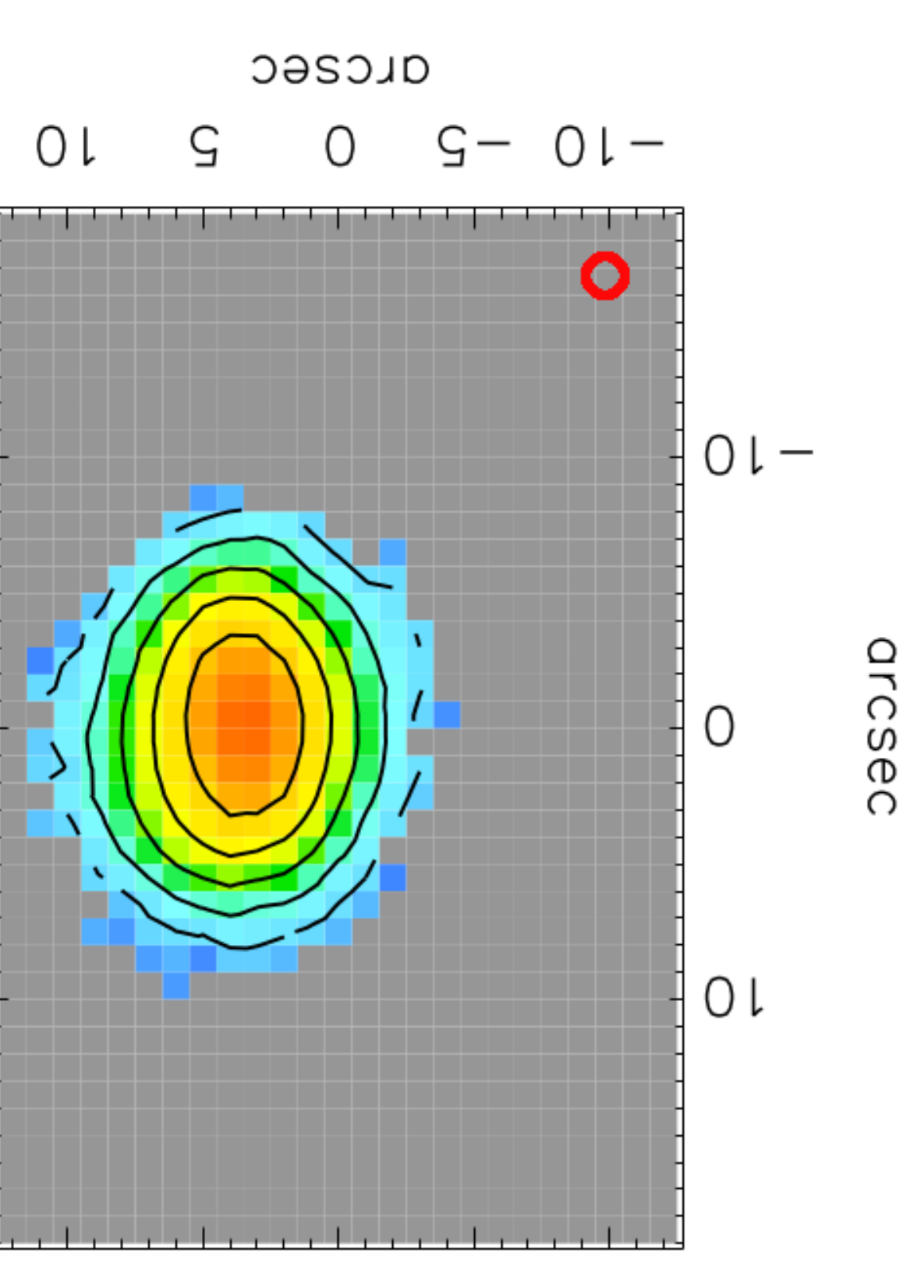} &
\includegraphics[keepaspectratio=true,height=33mm,width=25mm,angle=180,trim=127mm 28mm 32mm 20mm,clip=true]{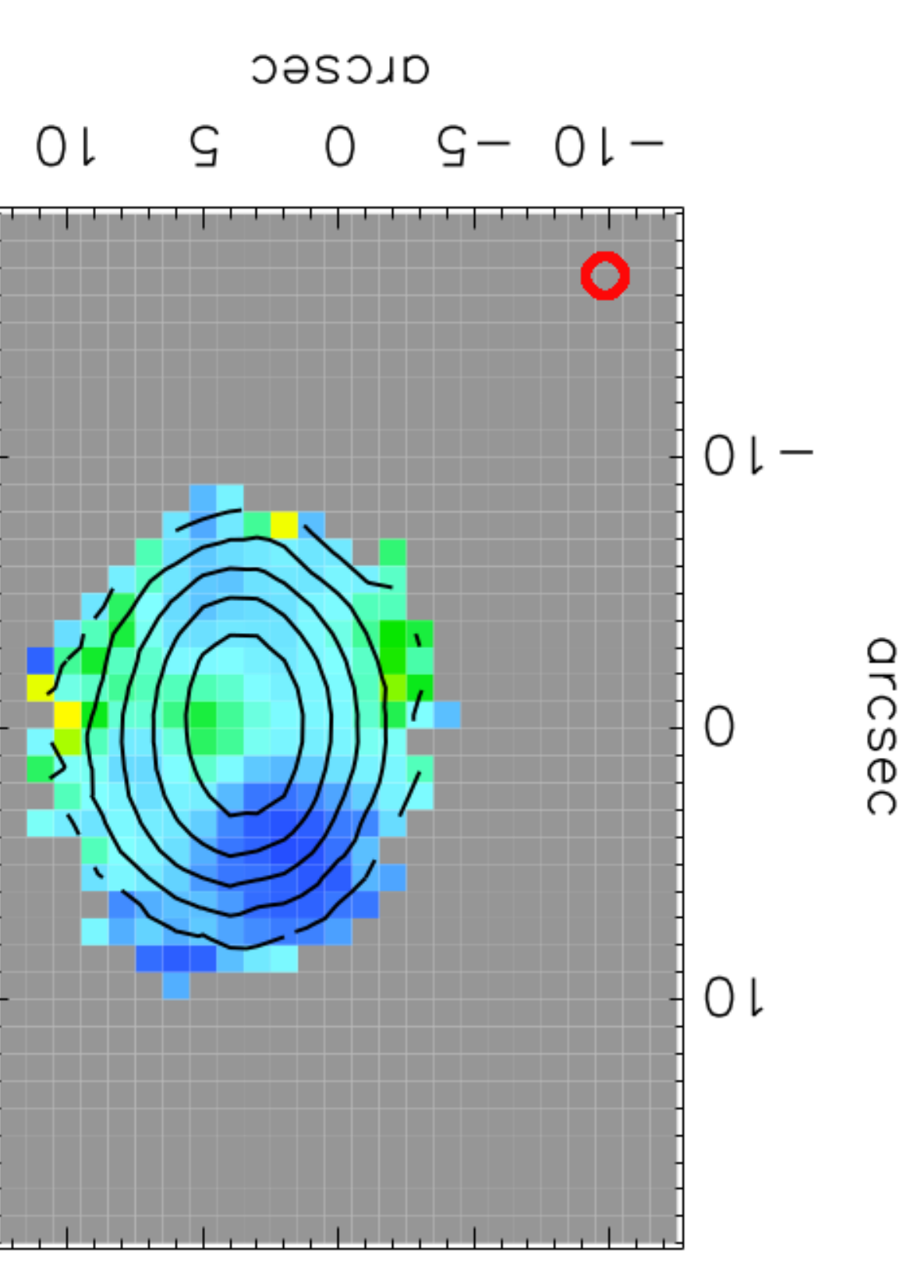} &
\includegraphics[keepaspectratio=true,height=33mm,width=25mm,angle=180,trim=127mm 28mm 32mm 20mm,clip=true]{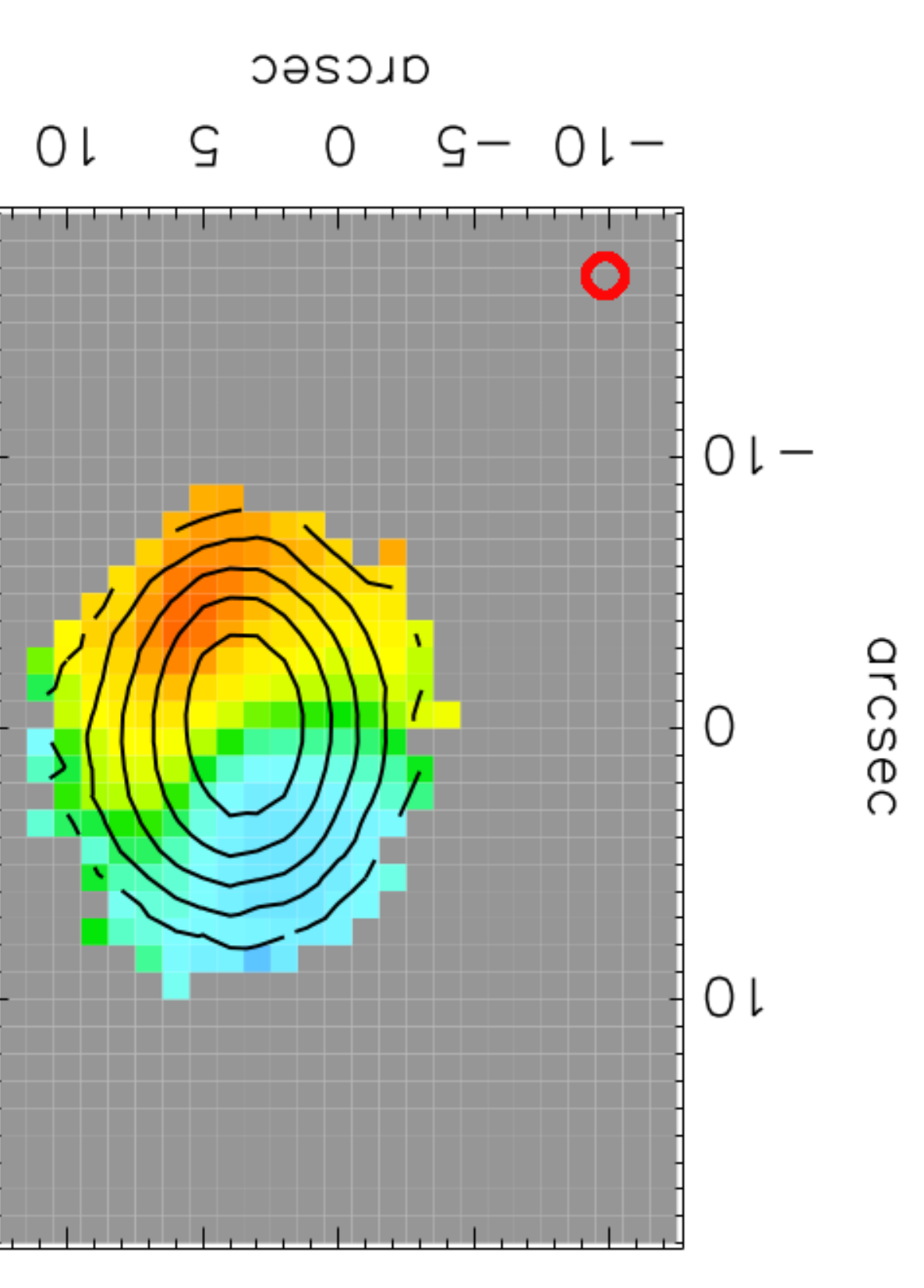} &
\includegraphics[keepaspectratio=true,height=33mm,width=25mm,angle=180,trim=127mm 28mm 32mm 20mm,clip=true]{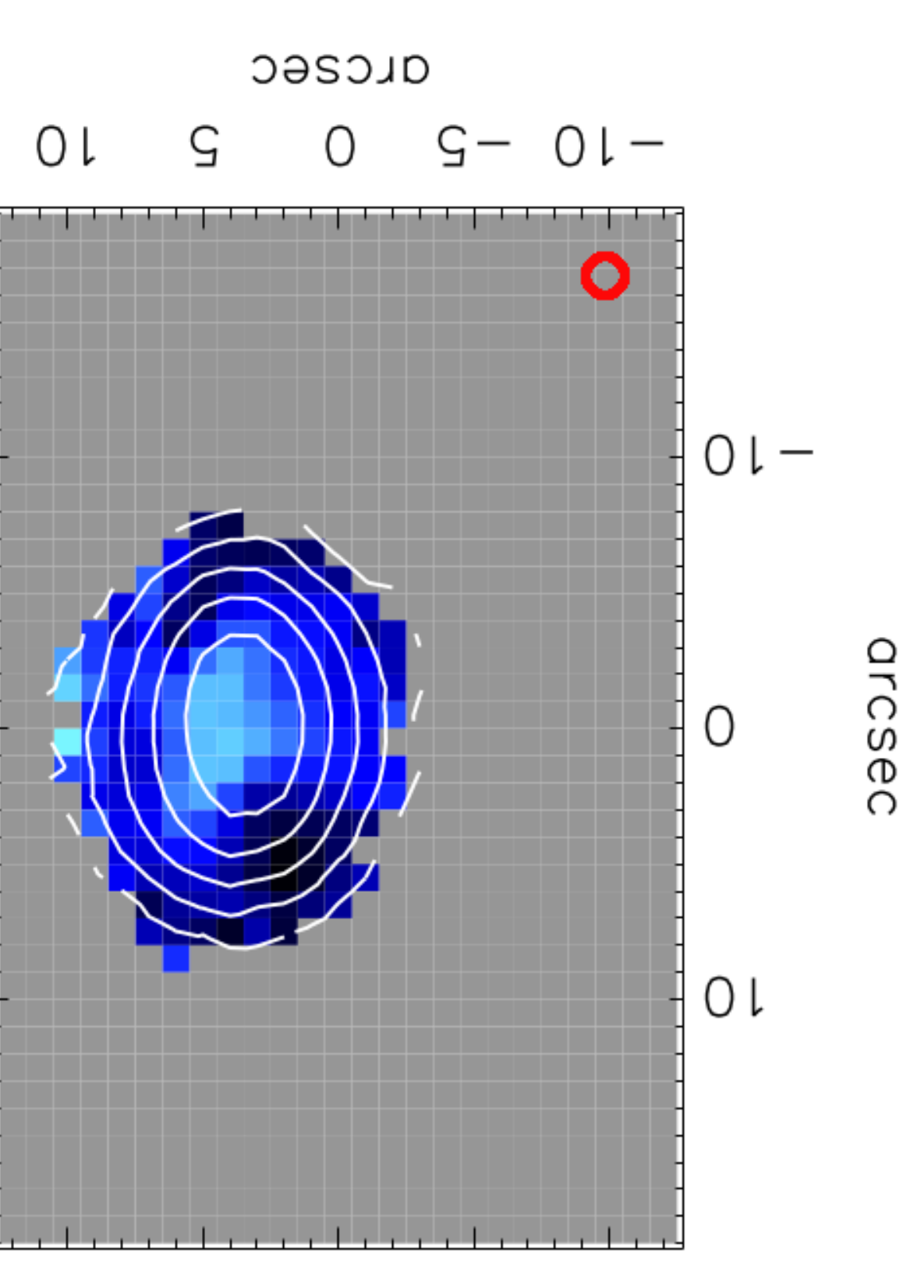} \\

F & \raisebox{\dimexpr-30.68mm+\ht\strutbox}{\includegraphics[keepaspectratio=true,height=27.1mm,width=27.5mm]{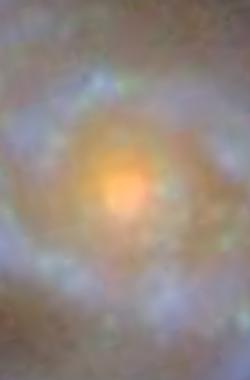}}&
\includegraphics[keepaspectratio=true,height=33mm,width=25mm,angle=180,trim=127mm 28mm 32mm 20mm,clip=true]{"gal7_p3b"} &
\includegraphics[keepaspectratio=true,height=33mm,width=25mm,angle=180,trim=127mm 28mm 32mm 20mm,clip=true]{"gal7_p2b"} &
\includegraphics[keepaspectratio=true,height=33mm,width=25mm,angle=180,trim=127mm 28mm 32mm 20mm,clip=true]{"gal7_p1b"} &
\includegraphics[keepaspectratio=true,height=33mm,width=25mm,angle=180,trim=127mm 28mm 32mm 20mm,clip=true]{"gal7_p4b"} \\

\end{tabular}
\caption{Kinematic properties of the observed galaxies. From left to right: thumbnail SDSS image; along with the line-of-sight measurements of the H$\alpha$ flux, velocity dispersion, velocity, and velocity gradient (left to right). The H$\alpha$ flux contours with $2.5\text{log}_{10}(F_{\text{H}\alpha})$  spacing are overlaid on all the kinematic maps. The size of the seeing disc (FWHM) 
is illustrated by the red circle in the bottom left corner of each map.}
\label{fig:kin}
\end{figure*}

We used the Wide Field Spectrograph \citep[WiFeS;][]{dopita2007,dopita2010} on the ANU 2.3-m telescope to obtain integral field spectroscopy of our sample. They were observed over two nights on the 10th and 11th of April 2013.
The total exposure time per object was between 1 and 2 hours and the seeing ranged between 1.4 and 2.0\,arcseconds. We used the R7000 grating which delivers a spectral resolution of $\sigma \sim 0.38\text{\AA}$ corresponding to a velocity  resolution of  $\sim 17$\,km\,s$^{-1}$ at H$\alpha$. The   wavelength range is  $5400\text{\AA} < \lambda < 7000\text{\AA}$ which includes the H$\alpha$ emission line at 6562.8\,$\text{\AA}$ and the [N II] emission lines at 6548.1\,$\text{\AA}$ and 6583.6,$\text{\AA}$. The WiFeS has a field-of-view of 25\,arcseconds $\times$ 38\,arcseconds, which covers most of the optical extent of our target galaxies (see the left most column of Figure \ref{fig:kin}). The spaxel size is 1\,arcsecond.

The data were reduced using the {\sc pywifes} data reduction package \citep{childress2014} which results in a fully reduced, co-added data cube. We performed a flux calibration using a matched aperture to the SDSS spectra.

\section{Data Analysis}
\label{sec:analysis}
\subsection{Emission line fits}
For each spaxel a triple Gaussian was fitted \citep[using MPFIT;][]{markwardt2009} to the H$\alpha$ emission line and  the [NII] emission lines at 6548.1 $\text{\AA}$ and 6583.6 $\text{\AA}$. The lines were fitted with a single velocity and velocity dispersion. In addition, the 6548.1\,$\text{\AA}$ [NII] emission line profile was fixed at the expected ratio of 1/3rd the amplitude of the [NII] emission line at 6583.6\,$\text{\AA}$.
As a result, the fit has four free parameters: the streaming velocity, velocity dispersion, and two line fluxes. The underlying continuum is taken into account by adding a linear term to the fit. 

We use the best fit parameters to construct parametric maps of the H$\alpha$ line flux, velocity dispersion and streaming velocity (see columns 2, 3 and 4 of Figure \ref{fig:kin}, respectively). 
The instrumental resolution, $\sigma_\text{r}$, was subtracted in quadrature from the observed velocity dispersion for each spaxel.
Low signal-to-noise spaxels were masked by requiring  the estimated H$\alpha$ amplitude to be greater than 2.5$\times$ the noise (standard deviation) in the continuum. Spaxels were also masked if the provided fits to the emission lines were poor. This was done by removing spaxels where the ratio of the fitted amplitude to the  actual H$\alpha$ peak  was less than 0.6. In addition, a limited number of individual, scattered spaxels in the outer regions were masked manually. The result of the masking can be seen in Figure \ref{fig:kin}.

\subsection{Star formation rates}
To measure the integrated star-formation of each galaxy we first
 co-added the spectra over the entire field-of-view of the IFU and
 measured the line fluxes of the H$\alpha$ and H$\beta$ lines.
The line fluxes were then corrected for stellar absorption following the prescription of
 \citet{hopkins2003}:
\begin{equation}
S = {EW + EW_{c} \over EW} F,
\end{equation}
where $S$ is the corrected line flux of the relevant line (H$\alpha$ or H$\beta$), EW is the equivalent width of the
line, EW$_{c}$ is the correction for underlying stellar absorption, and $F$ is the observed line flux. Again, following
\citet{hopkins2003} we assume EW$_{c}$=1.3\AA. Next we correct the H$\alpha$ line flux for dust extinction assuming
the dust extinction law of \citet{cardelli1989}:
\begin{equation}
F(H\alpha),\rm{corrected} =  F(H\alpha) (BD/2.86)^{2.36},
\end{equation}
where $BD$ is the Balmer Decrement: $F(H\alpha)/F(H\beta)$. The H$\alpha$ flux in then converted to a luminosity and
the star formation rate estimated using the \citet{kennicutt1998} relation and assuming a \citet{chabrier2003} IMF:
\begin{equation}
\rm{SFR} = {L(H\alpha) (W) \over 2.16\times 10^{34} } M\odot yr^{-1}.
\end{equation}
The star formation rates integrated over the IFU field-of-view are tabulated in Table \ref{tab:targets}.

\subsection{Beam smearing and velocity gradients}
Beam smearing increases the line-of-sight velocity dispersion since it mixes emission at different spatial locations, and hence different velocities, together. The magnitude of this effect depends on the local velocity gradient, in the sense that the greater the velocity gradient the greater the contribution of beam smearing to the line-of-sight velocity dispersion. 
We estimate the magnitude of the velocity gradient for a given spaxel $v_{g}(x,y)$  as the magnitude of the vector sum of the difference in the velocities in the adjacent spaxels, that is:
\begin{equation}
v_{\text{g}}(x,y) = 
\sqrt{
\begin{aligned}
& [v(x+1,y) - v(x-1,y)]^2 \\ 
 & \hspace{11mm} + [v(x,y+1) - v(x,y-1)]^2
\end{aligned}
}
\label{grad}
\end{equation}
Using this method, the velocity gradient is not defined for the spaxels on the edge of the field-of-view. Similarly, the velocity gradient is undefined if any of the adjacent spaxels were masked.
The velocity gradient maps are shown in the right most column of Figure \ref{fig:kin}.

\section{Results and Discussion}
\label{sec:results}
\subsection{Morphology in the parametric maps}
In the left-most column of Figure \ref{fig:kin} we show SDSS images of our sample. The morphologies are spiral galaxies with significant discs. 
In Figure \ref{fig:kin} we compare the spatial distribution of H$\alpha$ (column 2), the velocity dispersion (column 3), the streaming velocity (column 4) and the velocity gradient (column 5). 
For all galaxies, the line-of-sight H$\alpha$ flux and velocity gradient tend to peak in the central region of each galaxy. Similarly, the observed line-of-sight velocity dispersion tends to peak in the central region of each galaxy. The exception is galaxy E which does not show a clear peaked region of the line-of-sight velocity dispersion. The observed kinematics and magnitude of the velocity gradient are related to the inclination. Galaxies E and F are less inclined than the other galaxies based on their low ellipticity and shallow rotation curves. Those galaxies have lower observed velocity dispersion and velocity gradients in their centres.

Comparison of the parametric images provide insights into how  the observed velocity 
dispersion is effected by beam smearing and the local star formation rate. The velocity gradient map can be used as a proxy for where broadening of the emission lines due to beam smearing should be most prominent. Since convolution with the point-spread function mixes together a large range of velocities when the velocity gradient is large. While, the H$\alpha$ line flux can be used as a direct proxy for the local star-formation rate. 

It is apparent in Figure \ref{fig:kin} that the structure in the velocity dispersion maps is better matched to the structure in the velocity gradient maps in comparison to the structure in the H$\alpha$ emission line maps. For example, in Galaxy E, there is no central peak in the velocity dispersion, the velocity gradient is also relatively shallow in comparison to the other galaxies. Whereas the H$\alpha$ flux is still strongly peaked in the central region of the galaxy. 

For galaxies A, B, C, D and E the shape of the central velocity dispersion peak corresponds more closely to the shape of the peak in velocity gradient than it does with  the peak in H$\alpha$. Using galaxy B as an example, the relative H$\alpha$ flux is elongated in the south-west to north-east direction. However, the galaxy has a rotation axis, and thus a line-of-sight velocity gradient peak in the south-east to north-west direction. It can be similarly seen that the velocity dispersion has a peak region that is elongated in a direction that more closely matches that of the line-of-sight velocity gradient. 
Another example  are the non-central regions of strong H$\alpha$ flux in Galaxy A which correspond to regions of constant or decreased line-of-sight velocity dispersion. Those regions also have low local velocity gradients. The morphologies of the parametric maps suggest that local enhancements in the velocity dispersion correlate more closely with the local velocity gradient than the local star formation rate, and are suggestive of a beam smearing origin.

\subsection{Statistical analysis}
In Figure \ref{fig:scatter} we plot on a per spaxel basis the measured velocity dispersion  versus H$\alpha$ flux (left column) and the measured velocity dispersion versus the velocity gradient ($V_{g}$, right column). 
Each row represents a different galaxy in the sample.  To more clearly illustrate the extent to which the velocity dispersion correlates with each in the $F(H\alpha)$--$\sigma$ plane we colour code each value into quartiles of the velocity gradient. While in the $V_{g}$--$\sigma$ plane we colour code each value into quartiles of the H$\alpha$ flux. The quartile values are given in the legend in each panel. 

For each galaxy we fitted a 2D linear model with the velocity dispersion as the dependent variable and the H$\alpha$ flux and the velocity gradient as the independent variables:
\begin{eqnarray}
\begin{aligned}
 &\hspace{-1.2cm}\sigma_i[F(\text{H}\alpha_i),v_{\text{g},i}] =  \\
 &                                                                         m_{\text{H} \alpha} \text{Log}_{10}[F(\text{H}\alpha_i)]  +  m_{v_{\text{g}}} v_{\text{g},i} + C 
\label{eqn:fit}
\end{aligned}
\end{eqnarray}
The results are shown in Figure \ref{fig:scatter} and Table \ref{tab:spear}. In Figure \ref{fig:scatter} the resulting fits are shown for each quartile where the colour coded value has been held constant at its mean value in that quartile. 
\begin{table*}
\begin{center}
\caption{}
\label{tab:spear}
\begin{tabular}{ c@{ } \x c@{ }\x c@{ }\x c@{ }\x c@{ }\x c@{ }\x c@{ }\x c@{ }\x c@{ }\x c@{ }\x c@{ }\x}
\hline\hline
Label & $\rho^{a}_{(\sigma,\text{H}\alpha | v_{\text{g}})}$ & $\text{p}^{b}_{(\sigma,\text{H}\alpha| v_{\text{g}})} $ & $\rho^{a}_{(\sigma,v_{\text{g}}|\text{H}\alpha)}$ & $\text{p}^{b}_{(\sigma,v_{\text{g}}|\text{H}\alpha)}$ & $m^{c}_ {\text{H} \alpha}$ & $m^{c}_{v_{\text{g}}}$ & $\sigm$           & $\sigmu$                & $\sigmreg$      & $\sigmureg$ \\
         &                                                                 &                                                                       &                                                                &                                                                    &          \tiny{ ${(\rm{km}\,\rm{s}^{-1} )} \over \log(10^{-17} {\rm erg}^{-1} \rm{cm}^{-2})$}                                           &                              &   \tiny{ (km\,s$^{-1}$) } &     \tiny{(km\,s$^{-1}$)}      & \tiny{(km\,s$^{-1}$)}  &       \tiny{(km\,s$^{-1}$)}   \\
\hline
A &0.53 & $\sim$ 0 & 0.68 & $\sim$ 0 & 24$\pm$1    & 0.63$\pm$0.03 & 73.8$\pm$0.9 & 43$\pm$16 & $45\pm1$ & $24\pm19$\\
B &0.26 & 1.2e-4 & 0.81 & $\sim$ 0 & 5$\pm$2          & 0.47$\pm$0.02 & 72.3$\pm$0.7 & 57$\pm$16 & $32\pm1$ & $29\pm17$ \\
C &-0.43 & 1.5e-4 & 0.83 & $\sim$ 0 & -26$\pm$5    & 0.72$\pm$0.05 & 103.0$\pm$1.0 & 93$\pm$14 & $23\pm7$ & $35\pm16$\\
D &0.20 & 3.9e-3 & 0.86 & $\sim$ 0 & 8$\pm$3          & 0.73$\pm$0.04 & 65.0$\pm$1.0 & 48$\pm$6 & $26\pm2$ & $21\pm12$\\
E &-0.34 & 8.4e-6 & 0.70 & $\sim$ 0 & -8$\pm$2      & 0.60$\pm$0.05 & 76.8$\pm$0.4 & 75$\pm$10 & $44\pm1$ & $50\pm16$\\
F &0.42 & $\sim$ 0 & 0.57 & $\sim$ 0 & 5.9$\pm$0.4 & 0.60$\pm$0.02 & 34.5$\pm$0.1 & 29$\pm$5 & $25.6\pm0.1$ & $22\pm5$\\
\hline\hline
\end{tabular}
\end{center}
\tabnote{\hspace{5.5mm}$^a$Spearman's rank correlation coefficient.}
\tabnote{\hspace{5.5mm}$^b$Two-sided p-value from the null hypothesis $\rho_{(A,X|Y)} =  0$.}
\tabnote{\hspace{5.5mm}$^c$Slope for each independent variable supplied by the multiple linear regression fits.}
\end{table*}
Only four of the six galaxies show a positive correlation between velocity dispersion and H$\alpha$ flux. Galaxy C and Galaxy E have a mild negative correlation, i.e. the velocity dispersion increases with decreasing H$\alpha$ flux. Galaxy B and Galaxy D display mild positive correlations, while Galaxy A and Galaxy F display a highly significant positive correlation. On the other hand, in all cases there is a significant positive correlation between the velocity dispersion and the velocity gradient (see Table \ref{tab:spear}). This implies  beam smearing has a significant effect on the observed velocity dispersion even in these nearby galaxies.

There is a clear deviation from linear behaviour for the highest H$\alpha$ fluxes in Galaxy F. At H$\alpha$ fluxes above $\sim 5.5\times 10^{-17}$\,ergs/s/cm$^{2}$ there is a sharp upturn in the
velocity dispersion with increasing H$\alpha$ flux which cannot be explained by increases in the velocity gradient. There is also a hint of such an upturn in Galaxy B at approximately the same flux density and at lower flux density (albeit with larger scatter) in Galaxy D.  This is possible evidence for a local correlation between star-formation rate and gas velocity dispersion at the highest  H$\alpha$ fluxes. It is worth noting that galaxy F is the most nearby,  has the smallest maximum velocity gradient and the smallest velocity dispersion (at low velocity gradient) of any galaxy. Yet, it has some of the highest H$\alpha$ flux measurements. It may be that at such high SFR, star-formation feedback processes have a significant effect on the velocity dispersion, and that we can only clearly observe it in this galaxy due to the lack of beam smearing.

\begin{figure*}
\centering
\begin{tabular}{ r c\x c\x }

A &
\includegraphics[keepaspectratio=true,height=90mm,width=90mm,angle=180,trim=18mm 20mm 27mm 23mm,clip=true]{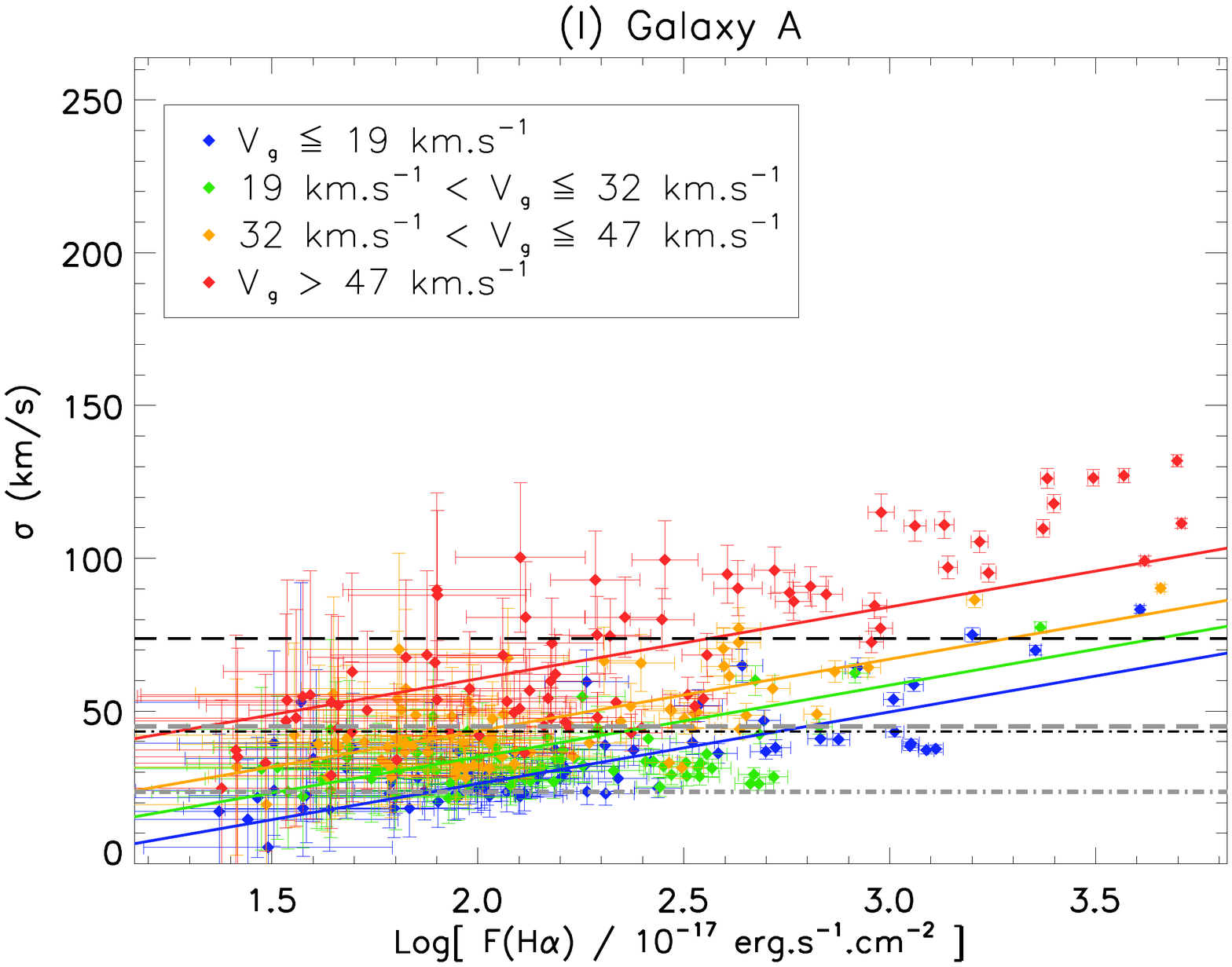} &
\includegraphics[keepaspectratio=true,height=90mm,width=90mm,angle=180,trim=18mm 20mm 27mm 23mm,clip=true]{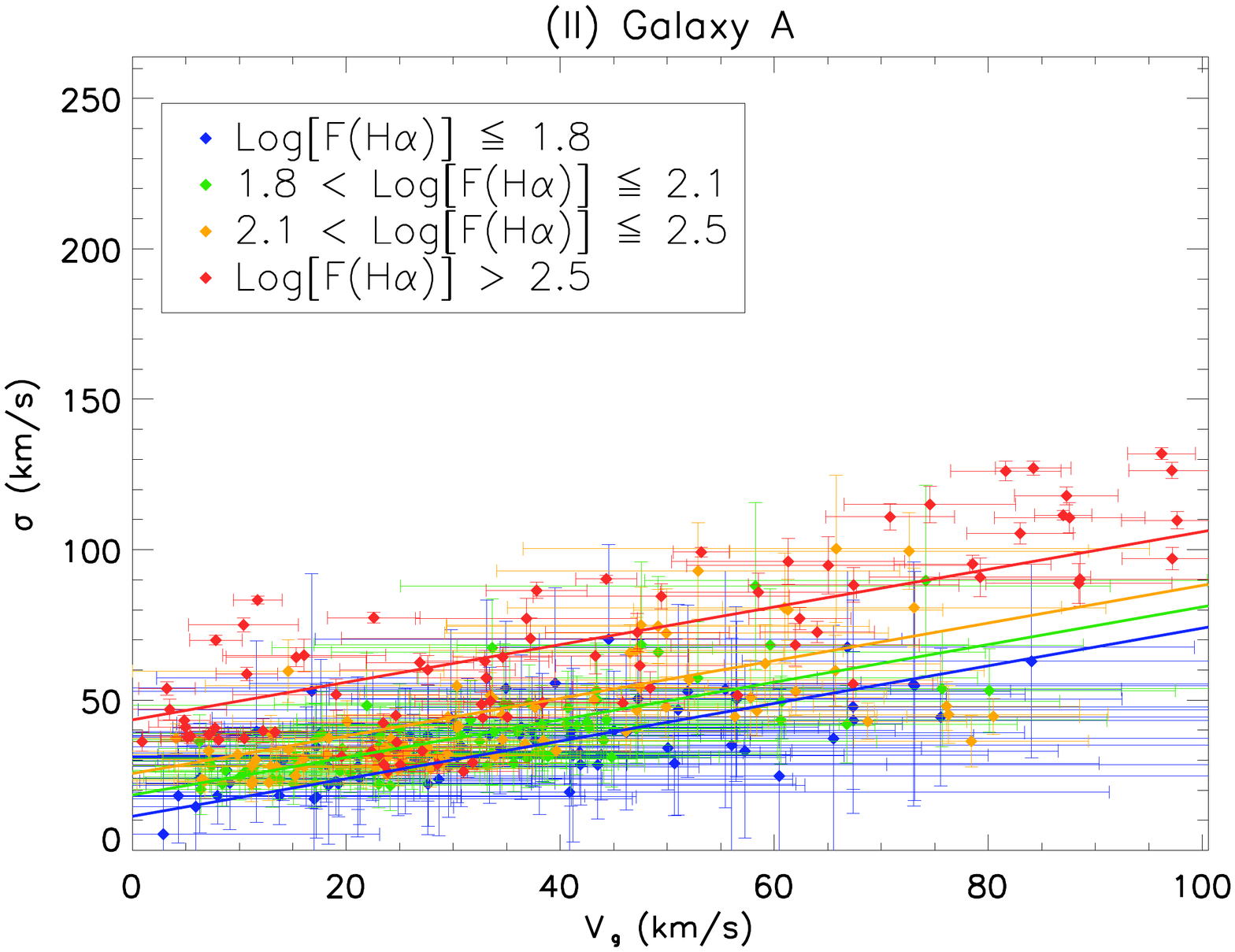} \\

B &
\includegraphics[keepaspectratio=true,height=90mm,width=90mm,angle=180,trim=18mm 20mm 27mm 23mm,clip=true]{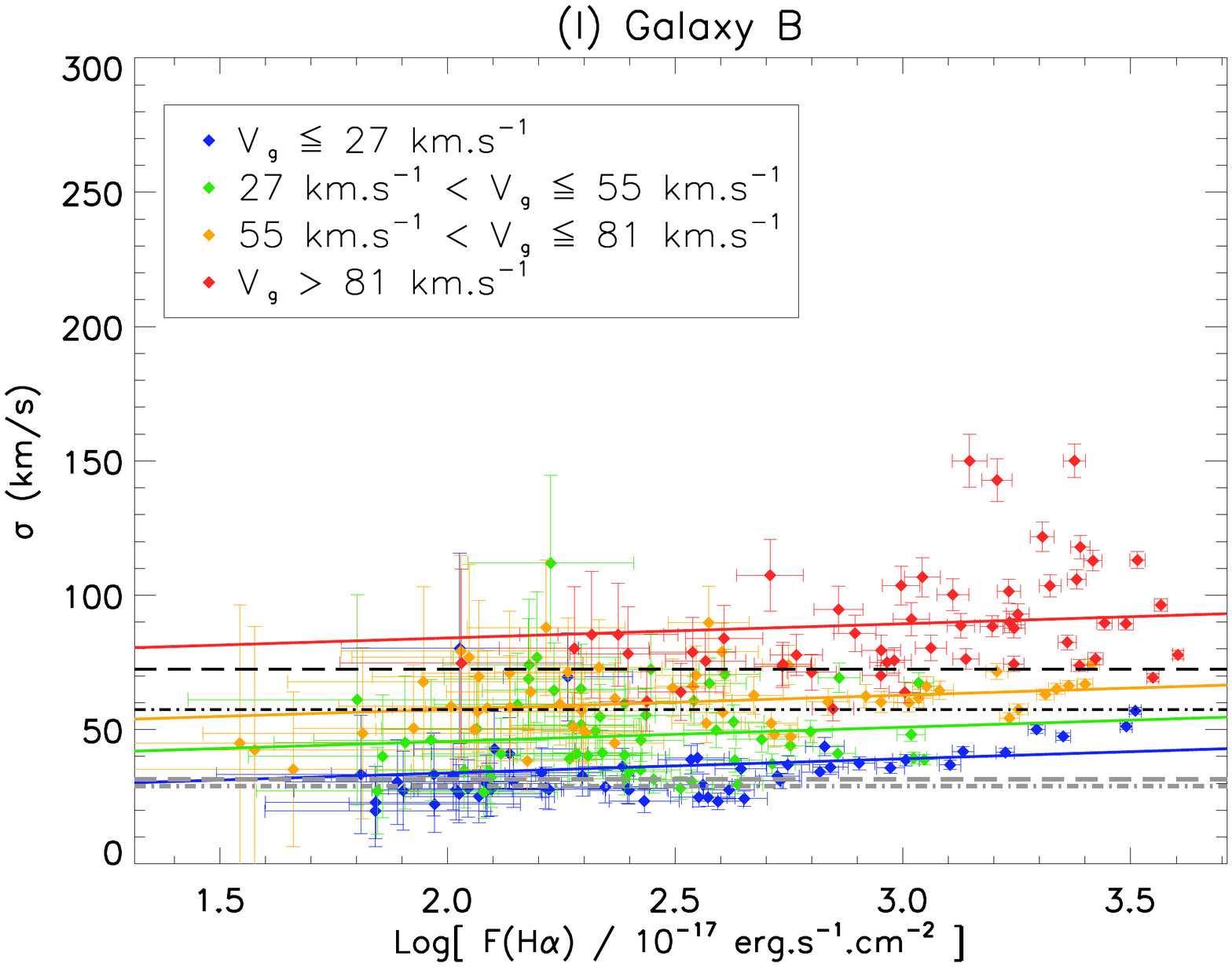} &
\includegraphics[keepaspectratio=true,height=90mm,width=90mm,angle=180,trim=18mm 20mm 27mm 23mm,clip=true]{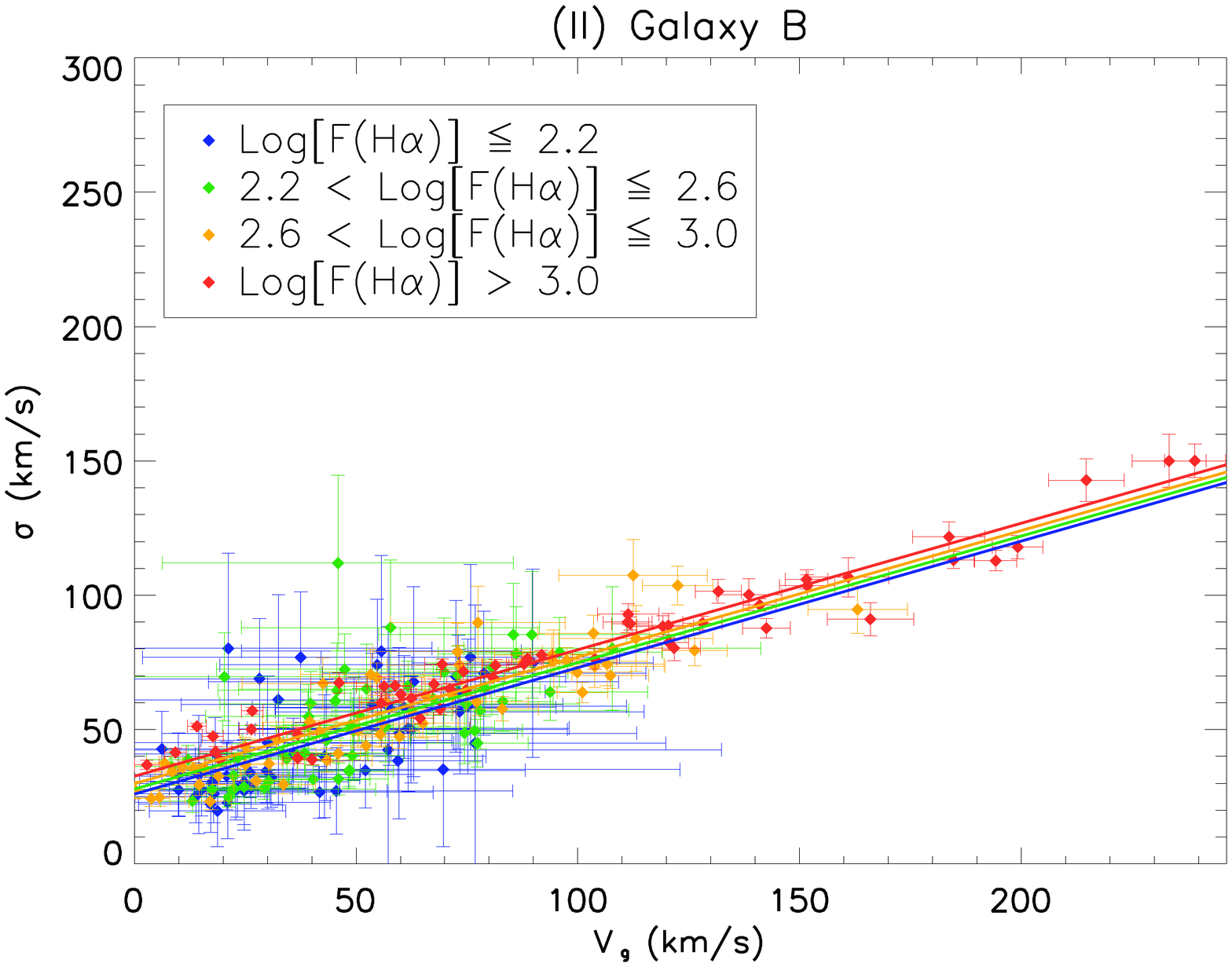} \\

C &
\includegraphics[keepaspectratio=true,height=90mm,width=90mm,angle=180,trim=18mm 20mm 27mm 23mm,clip=true]{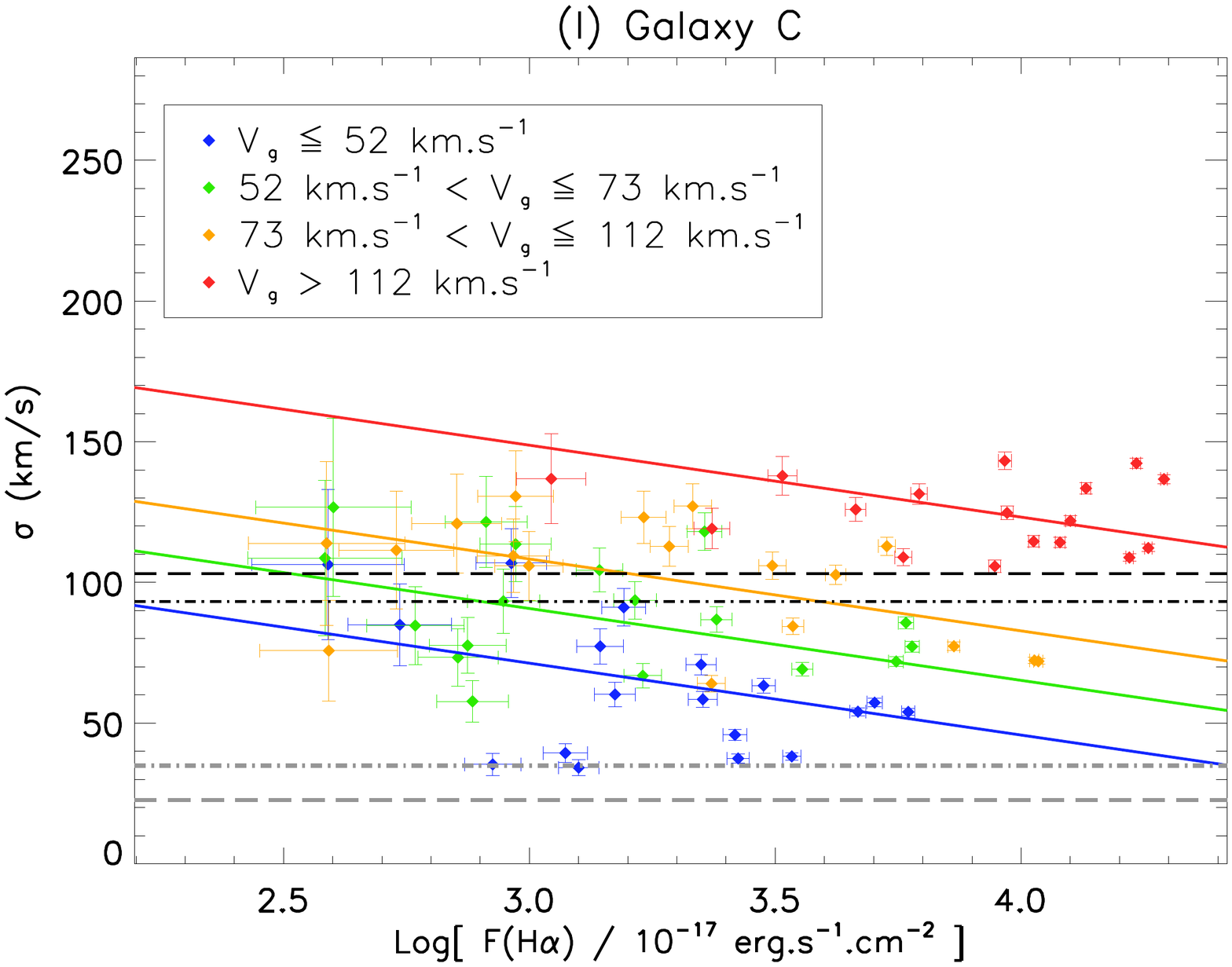} &
\includegraphics[keepaspectratio=true,height=90mm,width=90mm,angle=180,trim=18mm 20mm 27mm 23mm,clip=true]{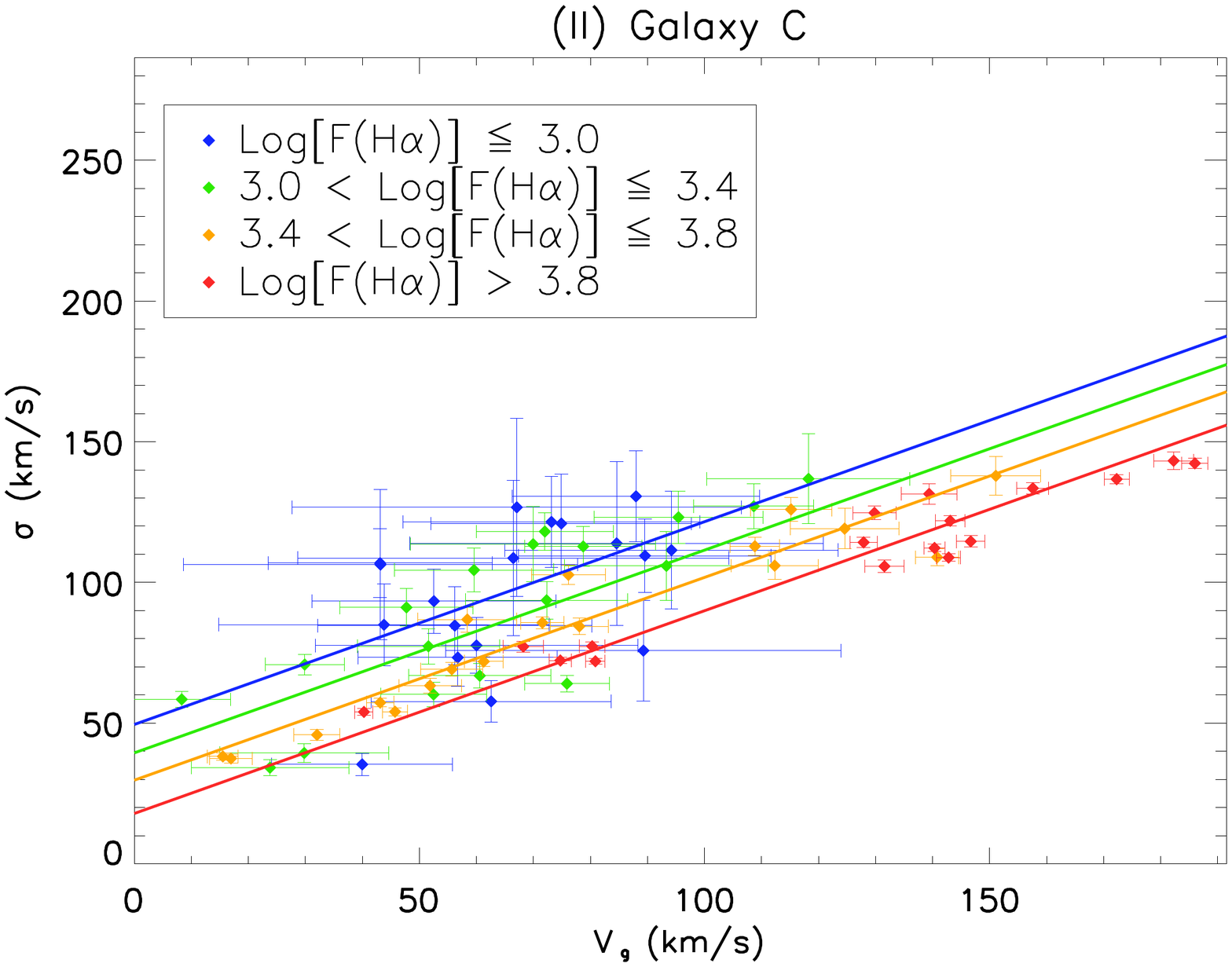} \\

\end{tabular}
\caption{Each row is for a seperate galaxy in the sample. Left column: The measured velocity dispersion  versus H$\alpha$ flux plotted for each spaxel. Right column: The measured velocity dispersion versus the velocity gradient ($V_{g}$). To  illustrate the effects of H$\alpha$ flux and  $V_{g}$ we colour code the spaxels into quartiles in the parameter not shown i.e. in the $F(H\alpha)$--$\sigma$ plane we colour code each value into quartiles of the velocity gradient and in the $V_{g}$--$\sigma$ plane we colour code each value into quartiles of the H$\alpha$ flux. The dashed and dot-dashed lines show the values of $\sigma_{m}$ and $\sigma_{m,uni}$, respectively. The uncorrected measurement is shown in black (higher values) and the corrected values (i.e.  $V_{g}=0$) in grey (lower values). }
\label{fig:scatter}
\end{figure*}

\setcounter{figure}{1}
\begin{figure*}
\centering
\begin{tabular}{ r c\x c\x }

D &
\includegraphics[keepaspectratio=true,height=90mm,width=90mm,angle=180,trim=18mm 20mm 27mm 23mm,clip=true]{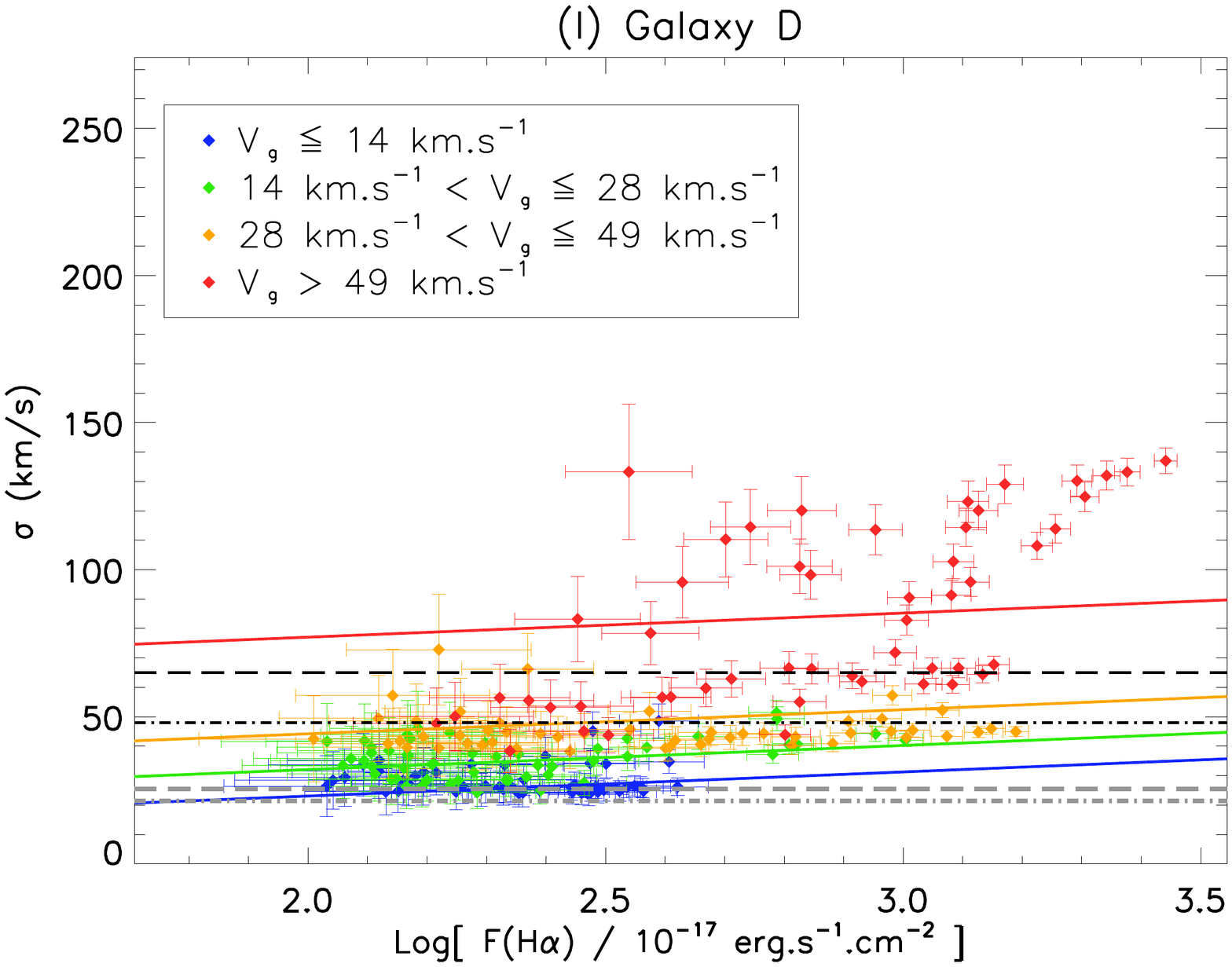} &
\includegraphics[keepaspectratio=true,height=90mm,width=90mm,angle=180,trim=18mm 20mm 27mm 23mm,clip=true]{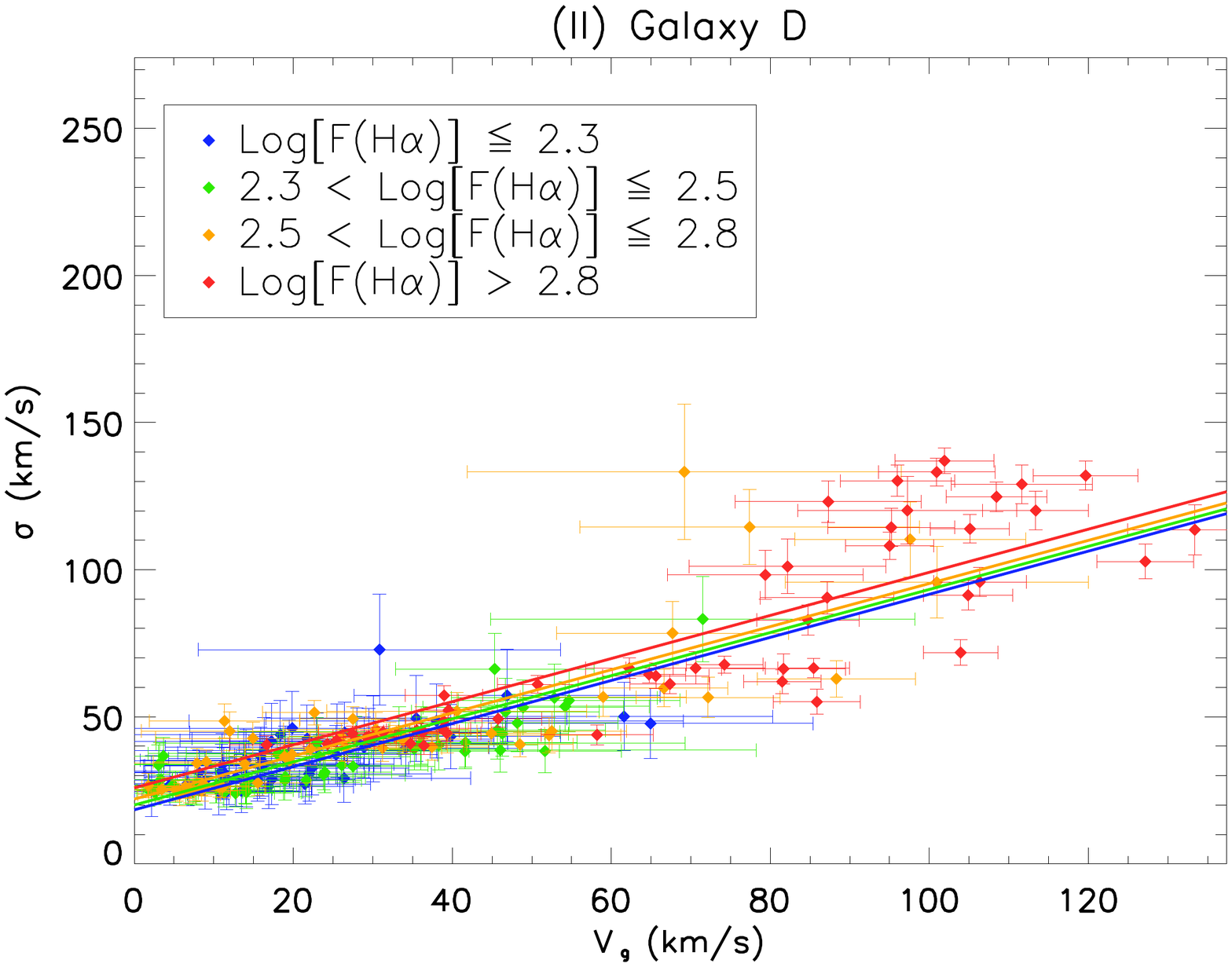} \\

E &
\includegraphics[keepaspectratio=true,height=90mm,width=90mm,angle=180,trim=18mm 20mm 27mm 23mm,clip=true]{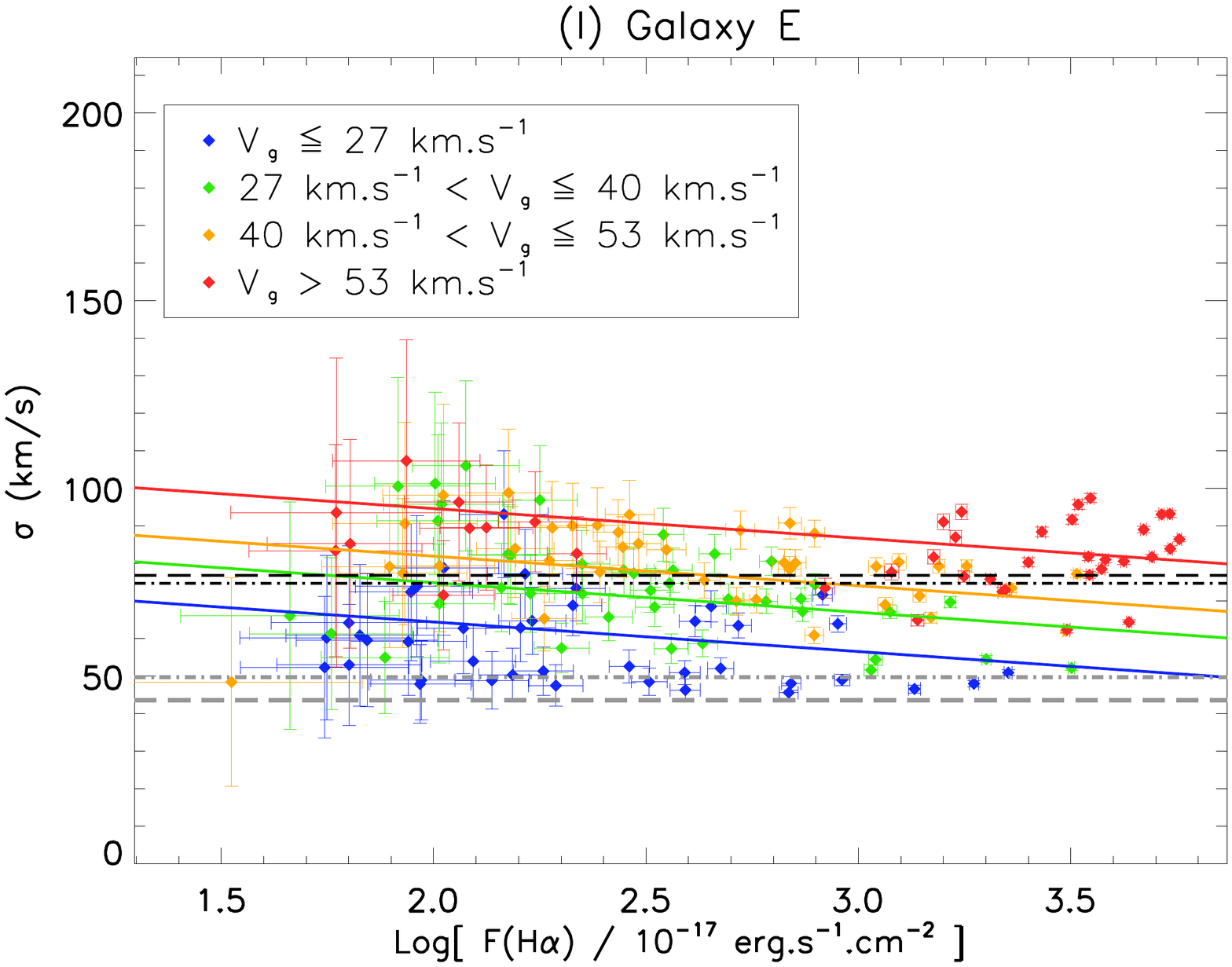} &
\includegraphics[keepaspectratio=true,height=90mm,width=90mm,angle=180,trim=18mm 20mm 27mm 23mm,clip=true]{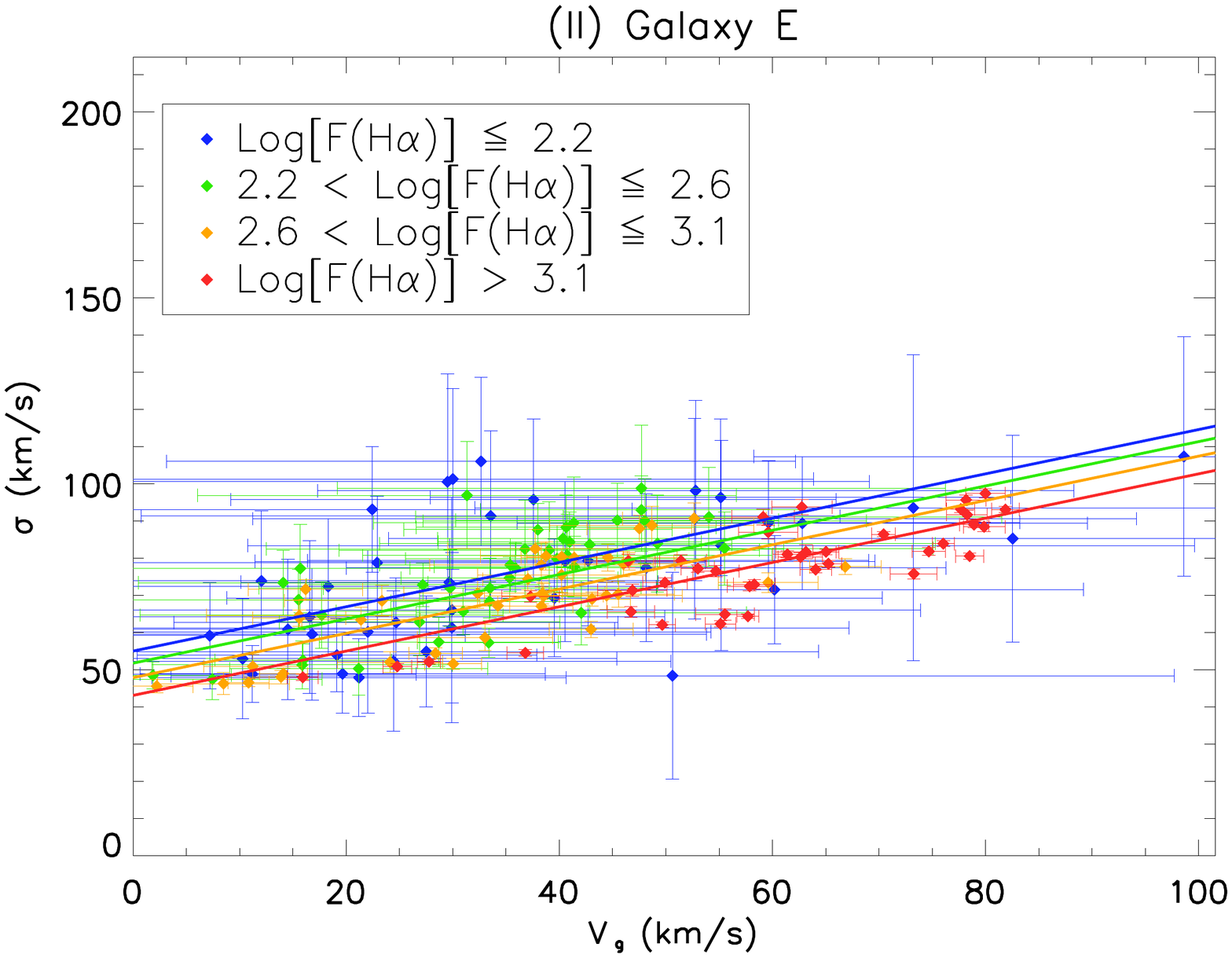} \\

F &
\includegraphics[keepaspectratio=true,height=90mm,width=90mm,angle=180,trim=18mm 20mm 27mm 23mm,clip=true]{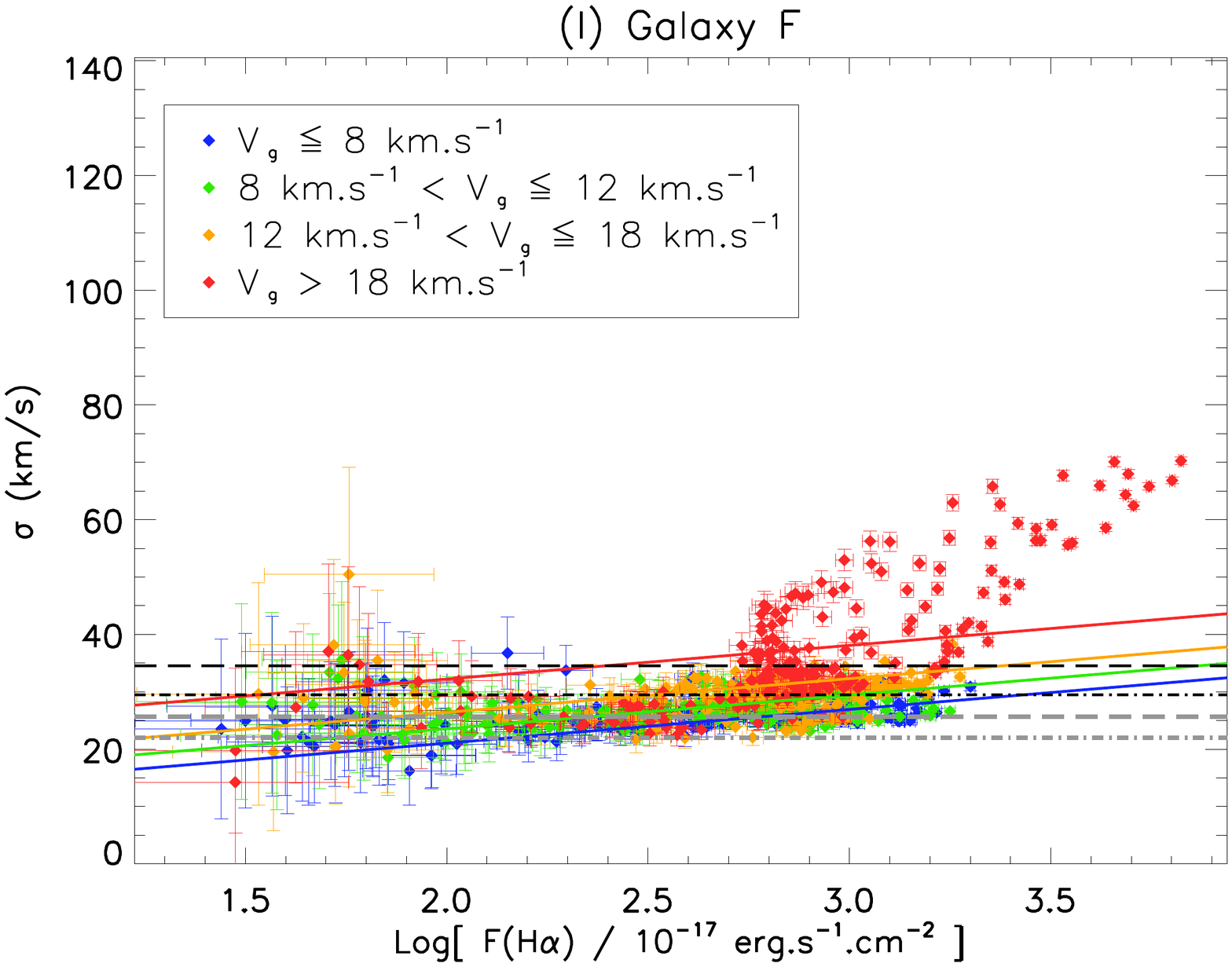} &
\includegraphics[keepaspectratio=true,height=90mm,width=90mm,angle=180,trim=18mm 20mm 27mm 23mm,clip=true]{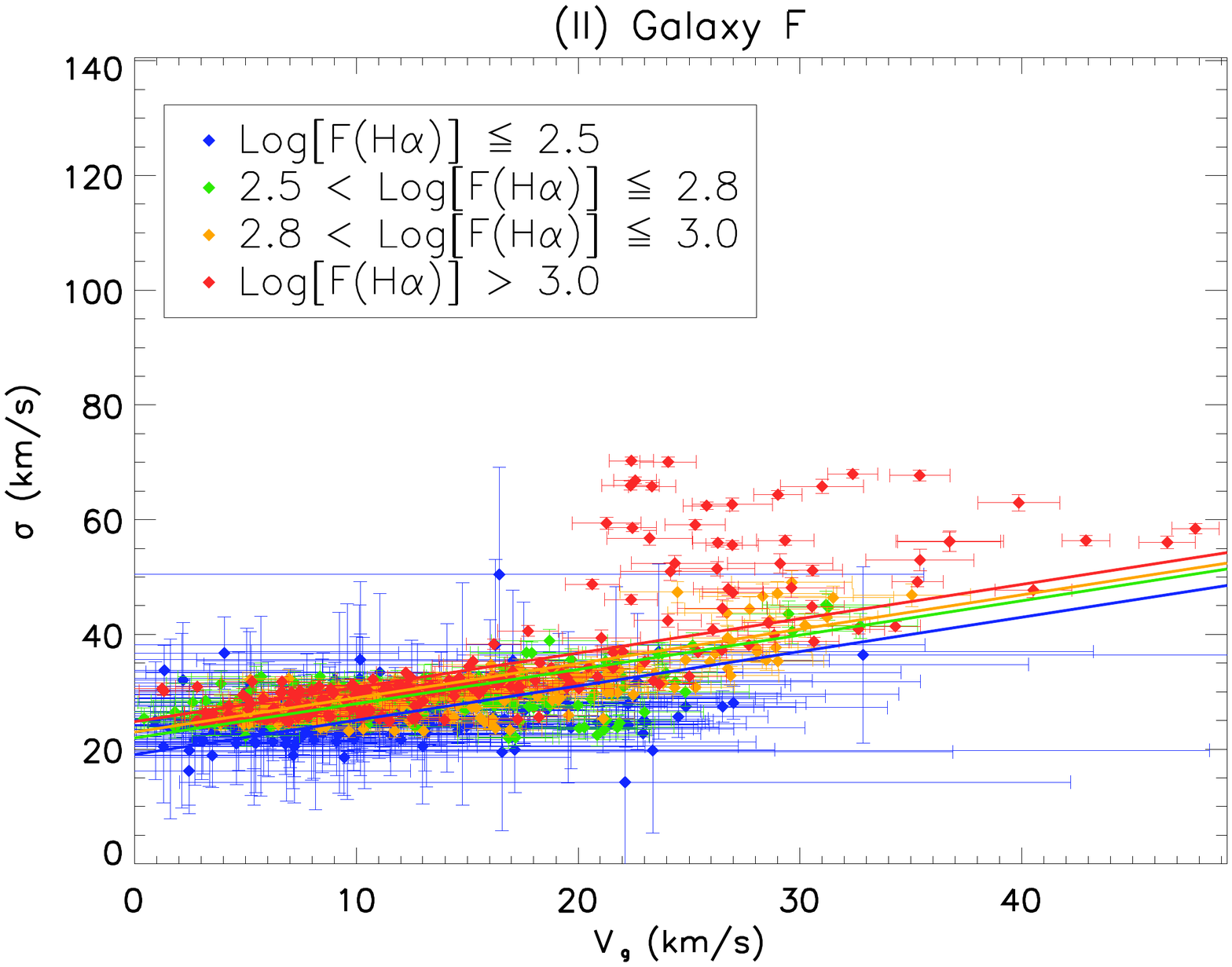}  \\

\end{tabular}
\caption{{\it cont.}}
\end{figure*}

The linear regression analysis above is only valid if the data are well represented by a linear model. While this is generally true there are deviations from this -- particularly at high H$\alpha$ flux. In
Table \ref{tab:spear} we also present statistics which do not require this assumption. Specifically, we calculate the the Spearman partial correlation coefficients defined as:
\begin{equation}
\rho_{(A,X|Y)} = \frac{\rho_{(A,X)} - \rho_{(X,Y)}\rho_{(A,Y)}}{\sqrt{(1 - \rho_{(X,Y)}^2)(1 - \rho^2_{(A,Y)})}},
\label{part}
\end{equation}
where this provides the correlation of A due to X given Y is held as a constant, along with the normalised two-sided p-value 
\begin{equation}
D_{(A,X|Y)} = \frac{ \sqrt{ N - 4 } }{ 2 }  ln \bigg( \frac{ 1 +  \rho_{(A,X|Y)}  }{ 1 - \rho_{(A,X|Y)} } \bigg),
\label{sig}
\end{equation}

These show a similar relationship without making the assumptions required using the multiple linear regression analysis.

\subsection{A simple non-parametric beam smearing correction}
The linear regression models in Figure \ref{fig:scatter} suggest a means for a simple non-parametric beam smearing correction. 
After fitting the linear regression model, the $\sigma$ value of each spaxel is taken and then corrected to $v_{\text{g},i}=0$  based on the best fit model . i.e. extrapolated to it's value at zero velocity gradient assuming the linear model.
This method  avoids assumptions about the kinematics of the line-of-sight velocity and 
velocity dispersion fields and is much simpler than disc model fitting. Although, it does require that the data are well represented by the linear model. This correction will
also be a conservative estimate since the velocity gradients are calculated from beam smeared data and the real velocity gradients will be larger.

\subsection{Global velocity dispersion}

The spatially resolved measurements of the velocity dispersion obtained using IFU spectroscopy can be combined into a single global velocity dispersion measurement. While reducing the parametric maps to a single value  decreases the information content, it is useful in order to make comparisons between different galaxies. However, there is no unique method of constructing a single velocity dispersion value from a two-dimensional map and as a result various measures have been used.  Popular methods, and the ones that will be explored in this paper,  include the flux-weighted mean $\sigm$:
\begin{equation}
\sigma_{m}={\sum\limits_{i=1}^{n} f_{i} \sigma_{i} \over \sum_{i=1}^{n} f_{i} }
\label{eqn:flux_weight}
\end{equation}
and a uniformly weighted mean denoted as  $\sigmu$:
\begin{equation}
\sigma_{m,uni}={\sum\limits_{i=1}^{n} \sigma_{i} \over n }
\label{eqn:arith_weight}
\end{equation}
Advantages of the flux weighted mean is that it up-weights high signal-to-noise spaxels; it is not sensitive to how low signal-to-noise regions in the galaxy (usually the outskirts) are masked; and since the flux of the H$\alpha$ line is proportional to the star formation rate it is the natural way to combine the data when this is the process of interest. One disadvantage is that the star-formation rate is often greatest in the central region of the galaxy and this is also where beam smearing issues are maximal. This results in a bias towards higher global velocity dispersion \citep[e.g.][]{davies2011}.
\begin{figure}
\label{sec:green_beam}. 
\includegraphics[width=90mm]{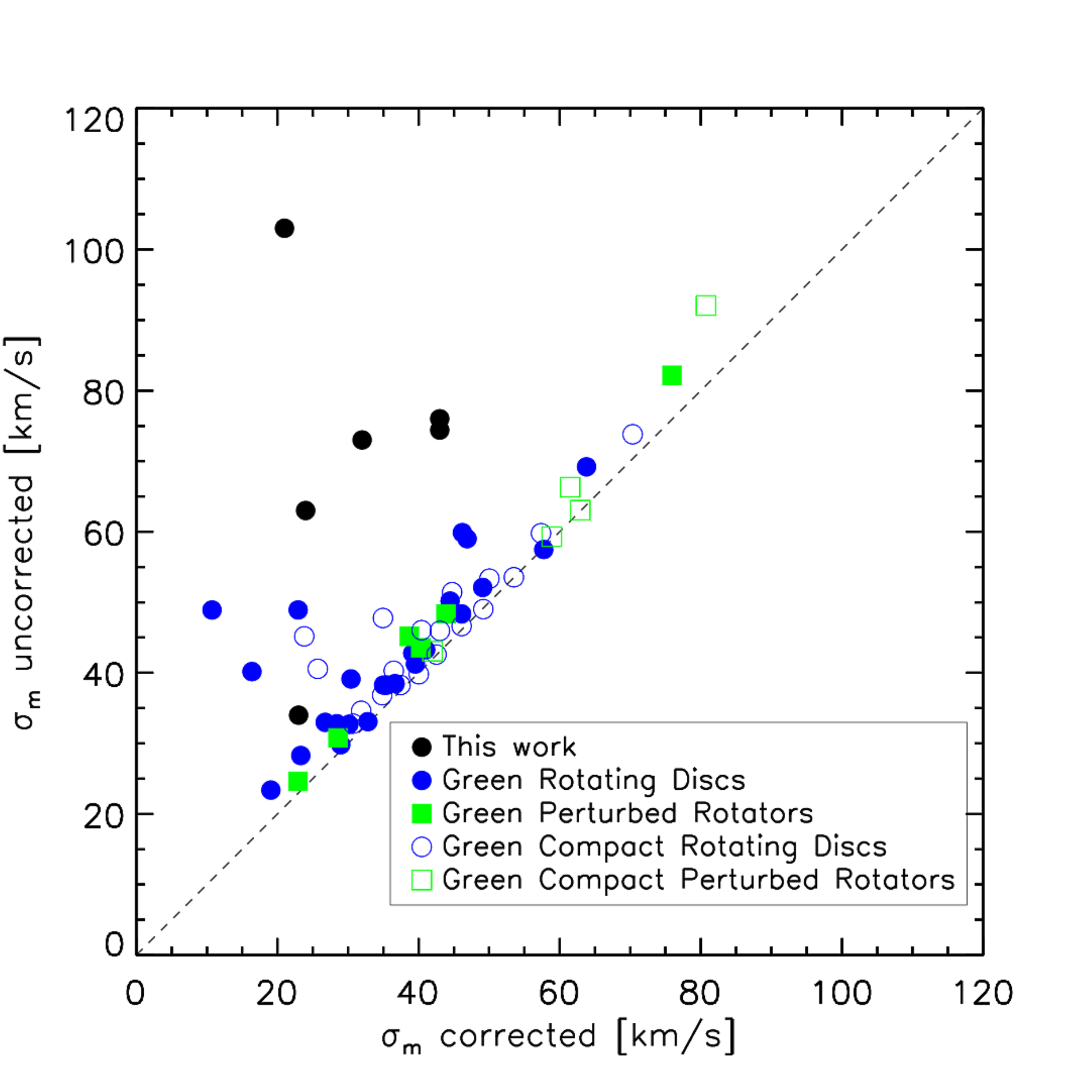}
\caption{The uncorrected values of $\sigma_m$ compared to our beam smearing corrected values (black filled circles). Also shown are the sample of rotating galaxies from \citet{green2014} where the beam smearing correction is done using disc model fitting.}
\label{fig:beam_correction}
\end{figure}

This bias can be mitigated to some extent by taking a uniformly weighted mean. However, $\sigmu$ can be greatly affected by the imposed masking techniques.  Outskirts of the galaxy will tend to have lower signal/noise affecting the accuracy in the derived kinematics. The beam smearing bias in the central region is still present and its effect increases as the signal-to-noise constraints used in masking are tightened.

In Table \ref{tab:spear} we show our derived global velocity dispersion calculated on a flux-weighted basis using equation \ref{eqn:flux_weight} and a straight arithmetic mean using equation \ref{eqn:arith_weight}.   In every case the flux-weighted mean velocity dispersion $\sigma_{m}$ is larger than the arithmetic mean $\sigma_{m,uni}$. The reason for this is evident from the maps in Figure \ref{fig:kin}. The H$\alpha$ flux is generally centrally concentrated and this is also where the velocity field is steepest and the effect of beam smearing is maximised. Equation \ref{eqn:flux_weight} up-weights these region producing a larger value of the velocity dispersion. We also show our values of $\sigma_m$  and $\sigma_{m,uni}$  after applying our simple beam smearing correction, calculated using the best fitting solution to Equation \ref{eqn:fit} and setting  $v_{\text{g},i}=0$. The corrected values  are $\sim 1.5 - 4.5$ and $\sim 1.3 - 2.7$ lower  than the  uncorrected values of $\sigma_{m}$ and $\sigma_{m,uni}$ respectively.

In Figure \ref{fig:beam_correction} we plot our raw versus corrected values of $\sigma_{m}$. The points always lie above the one-to-one line since the beam smearing correction should (and does) only decrease the
value of $\sigma_{m}$. For comparison we show the sample of rotating galaxies from \citet{green2014}. These values have been beam smearing corrected using a disc model and the assumption of an exponential light profile to calculate (on a per spaxel basis) the contribution to the velocity width arising from the unresolved velocity gradient across the spaxel. This `beam smearing map' is subtracted in quadrature from the observed velocity dispersion map to produce a corrected velocity dispersion map. The velocity dispersion value is then calculated using equation \ref{eqn:flux_weight}. The magnitude of the corrections are generally much smaller than those found using our simple linear extrapolation to zero-velocity-gradient and many of the  \citet{green2014} galaxies lie close to the one-to-one line. The median correction being just $3.6$\,km\,s$^{-1}$.

\subsection{Global velocity dispersion and star formation rate}
\begin{figure*}
\begin{center}
\includegraphics[height=100mm,width=130mm,angle=180,trim=18mm 21mm 25mm 23mm,clip=true]{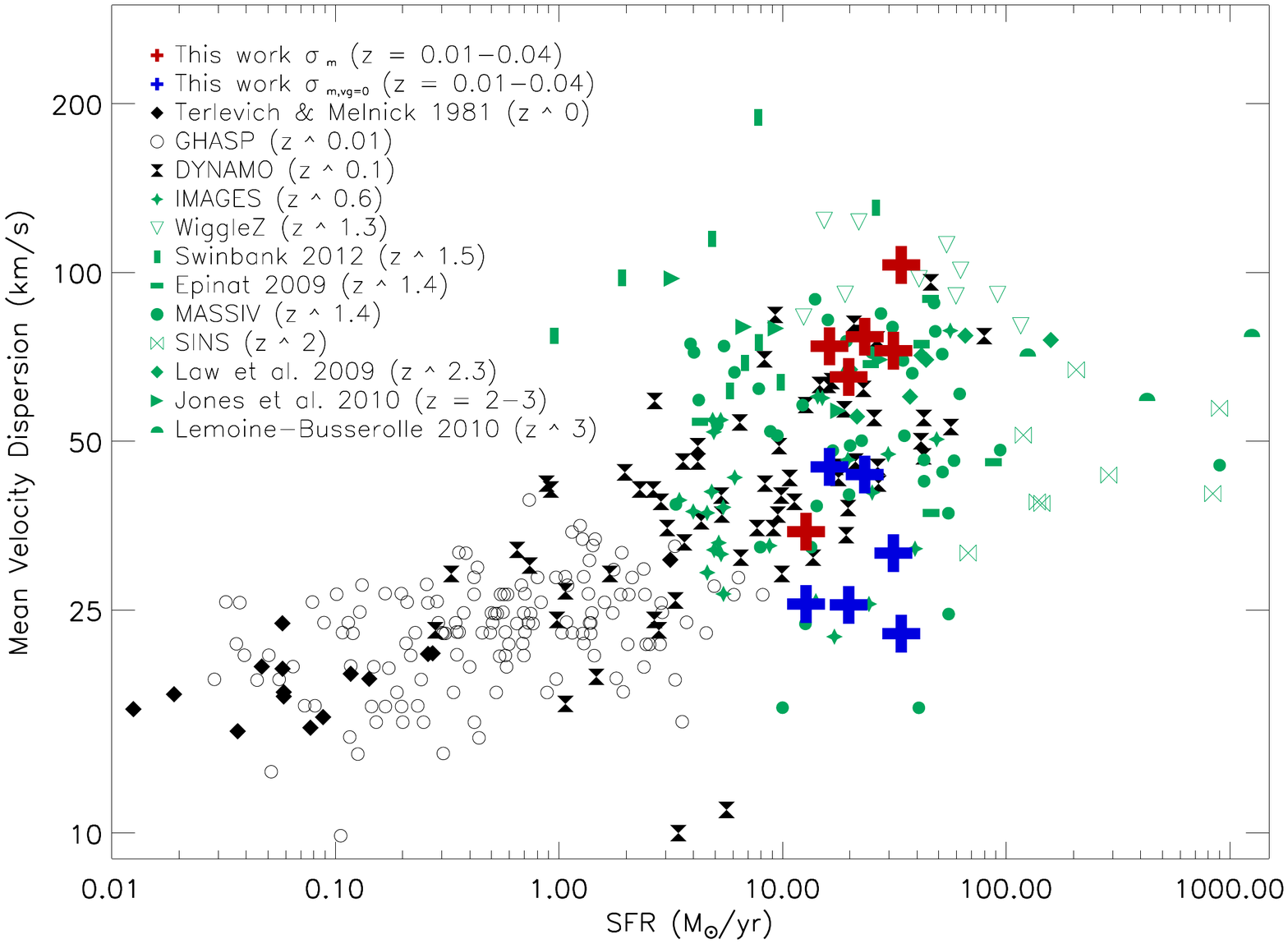}
\caption{The star formation rate plotted against mean velocity dispersion. Our flux weighted values ($\sigma_m$)  are shown as red crosses. The beam smearing corrected flux weighted values ($\sigma_{m,v_g=0}$) are shown as  blue crosses. A compilation of literature values are also shown with their SFRs converted to the assumption of a  \citet{chabrier2003} IMF.  Black symbols are for samples at $z<0.5$ and green symbols are for objects at $z>0.5$. The sample of \citet{terlevich1981} is displayed as filled black diamonds. Measurements from the GHASP survey are shown as open black circles \citep{epinat2008a,epinat2008b,garrido2002}. The DYNAMO sample is shown as filled black hourglass symbols \citep{green2014}. Results from the IMAGES survey \citep{yang2008} are shown as filled green stars. Objects selected from the WiggleZ survey are shown as open upside-down triangles \citep{wisnioski2011}.  The samples of \citet{swinbank2012} and \citet{epinat2009} are displayed as vertical filled green rectangles and horizontal filled green rectangles, respectively. Measurements from the MASSIV survey \citep{contini2012}  and the SINS survey \citep{cresci2009} are shown as filled green circles and open bow-tie symbols, respectively.  The samples of \citet{law2009}, \citet{jones2010} and \citet{lemoine-busserolle2010} are displayed as filled green diamonds, triangles and semi-circles, respectively. Note:  the $\sigma$ values measured by \citet{terlevich1981} are integrated over small sub-galactic scales of $\sim$10--100pc.
}
\label{fig:compare}
\end{center}
\end{figure*}

There is an observed  positive correlation between $\sigma_{m}$ and the star formation rate which appears independent of redshift \citep{green2010,green2014}, and this correlation has been used to argue that star-formation is driving the turbulence in galaxies at all cosmic epochs. In Figure \ref{fig:compare} we show our derived global velocity dispersion calculated on a flux-weighted basis ($\sigma_{m}$) plotted against the  global star formation rate (red crosses) calculated from summing over the entire IFU field-of-view (assumes a \citet{chabrier2003} IMF). We also show our values of  $\sigma_{m}$ after applying our simple beam smearing correction (blue crosses). 

Also shown in Figure \ref{fig:compare} is a compilation of results from the literature taken from \citet{green2014}. The star formation rates have all been corrected for dust extinction and \citet{green2014} converted each to their adopted \citet{chabrier2003} Initial Mass Function. However, the star-formation rates are derived from different indicators (e.g. H$\alpha$, SED fitting, UV luminosity) and the methods used in determining the velocity dispersions also vary.   

 At the lowest redshifts ($z \lesssim 0.01$ ) are  observations of HII regions by  \citet[][]{terlevich1981} where the velocity dispersion is a spatially integrated measurement over scales of 10--100pc, and observations of disc galaxies in the GHASP survey \cite[][]{epinat2008a,epinat2008b,epinat2010,garrido2002} which uses an unweighted mean (i.e. $\sigma_{m,uni})$. At intermediate redshift ($0.4<z<0.75$) we show the results from the IMAGES survey \citep[][]{yang2008} again using the unweighted mean. At redshifts between  $z=1$ and $z=2 $ we show the MASSIV survey \citep{contini2012} and the samples of  \cite{wisnioski2011} and  \cite{swinbank2012}; which use a flux-weighted mean velocity dispersion. At $z\sim $2--3 we show the SINS survey which uses a velocity dispersion derived from disc modelling \citep{cresci2009} which has the effect of up-weighting the outer, least beam smeared, parts of the disc. Also at   $z\sim $2--3 is the sample  of \citet{epinat2009} which has the average weighted by the inverse-error  and the gravitationally lensed sample of \citet{jones2010} as well as the samples of \citet{lemoine-busserolle2010} and \citet{law2009}; all of which used $\sigma_m$.

The DYNMO survey (shown as black hourglass symbols in Figure \ref{fig:compare}) of \citet[][uses $\sigma_{m}$]{green2014}  at $z\sim$0.055--0.3  was designed to bridge the gap between the high and low redshift surveys by including galaxies with high star-formation rates similar to those at high redshift but with much better physical-scale resolution and surface brightness limits. These values have been beam smearing corrected using a disc model and the assumption of an exponential light profile as described in Section \ref{sec:green_beam}. 

Our uncorrected beam smearing measurements of the mean velocity dispersion (red crosses) are in good agreement with the compilation of results from other surveys. They are consistent with the higher values of velocity dispersion at high star-formation rates and the scatter in the distribution is similar to the literature results; with values of $\sigma_m$ ranging from $\sim$25--100\,km\,s$^{-1}$. The beam smearing corrected measurements have smaller values by a factor of $\sim 2$ and range from $\sim$20--50\,km\,s$^{-1}$. These  only overlap with the lower end of the velocity dispersion range  (at similar star-formation rate) in the literature data and imply a shallower relationship between $\sigma_m$ and star-formation rate. 

Given we expect our beam smearing correction to, in general, underestimate the true correction since it uses the observed data to calculate the velocity gradient, our data alone do not necessarily imply a positive correlation between $\sigma_m$ and global star formation rate. As pointed out by \citet{green2014} because of the large variety in the data and analysis techniques contributing to Figure \ref{fig:compare} it is not appropriate to quantify the slope of this relation. The large difference in the slope that is implied by our corrected and uncorrected measurements and the significant scatter support the need for a more uniform set of measurements and a credible and accurate account of beam smearing across the entire parameter space.

\section{Summary}
We have used integral field spectroscopy of a sample of six nearby, $z \sim 0.01 - 0.04$, high star-formation rate ($\text{SFR} \sim 10 - 40$ $\text{M}_\odot \text{/yr}$) galaxies to investigate the relationship between  velocity dispersion and star formation rate. The low redshift selection was made to minimise the effects of beam smearing which artificially increases the measured line-of-sight velocity dispersion by
including unresolved rotation into the measured dispersion. We found:
\begin{itemize}
\item The contribution of beam smearing to the velocity dispersion is still significant, even at these low redshift. This is evidenced by the strong correlation between the local (spaxel by spaxel) velocity gradient (a proxy for beam smearing) and the velocity dispersion. 

\item The correlation between velocity gradient and velocity dispersion is stronger than the correlation between H$\alpha$ flux (a proxy for star formation rate) and velocity dispersion. When measured on a spaxel-by-spaxel basis the former shows a positive correlation in all six galaxies while for the latter this is true in only 4 of 6 cases. 

\item We present a simple non parametric beam smearing correction based on a 2D regression model of the velocity dispersion in a spaxel with respect to the velocity gradient and H$\alpha$ flux and extrapolating to zero velocity gradient. This results in corrections to the velocity dispersion of  a factor of $\sim  1.3 - 4.5$ and $\sim 1.3 - 2.7$  for the uncorrected flux-weighted and unweighted mean line-of-sight velocity dispersion values,  respectively.

\item The beam smearing corrected values of  the mean velocity dispersion ($\sigma_m \sim 20 - 50$ km/s)  are only marginally larger than those found in nearby low star-formation rate galaxies  ($\sigma_m \sim 10 - 25$ km/s).

\end{itemize}

\begin{acknowledgements}
SMC acknowledges the support of an Australian Research Council Future
Fellowship (FT100100457). Parts of this research were conducted by the Australian
Research Council Centre of Excellence for All-sky  Astrophysics
(CAASTRO), through project number CE110001020.
We are grateful to Adam Schaefer for helpful discussions concerning this work.
We would also like to thank the anonymous referee for insightful
comments which greatly improved this paper.

\end{acknowledgements}

\bibliographystyle{apj}
\bibliography{references.bib}

\end{document}